\documentclass[aps,pra,nobalancelastpage,twocolumn,superscriptaddress,nofootinbib,amsmath,groupedaddress,longbibliography]{revtex4-1}

\usepackage{CJKutf8}
\usepackage{graphicx}
\usepackage{subfigure}
\usepackage{float}
\usepackage{dcolumn}
\usepackage{bm}
\usepackage{siunitx}
\usepackage{color}
\usepackage{longtable}
\usepackage{amsmath}
\usepackage{amssymb}
\usepackage{array}
\usepackage{multirow}
\usepackage{tablefootnote}
\usepackage{upgreek}
\usepackage{braket}
\usepackage{makecell}
\usepackage{physics}
\usepackage[letterpaper,total={7in,9.5in},top=0.75in,left=0.75in]{geometry}
\usepackage[colorlinks=true,linkcolor=blue,citecolor=blue,urlcolor=blue,filecolor=blue]{hyperref}
\usepackage{lipsum}  
\usepackage{xcolor}
\usepackage{ulem} 
\usepackage{orcidlink}
\usepackage{titlesec}
\titlespacing*{\section}{0pt}{5pt}{5pt}

\definecolor{lime}{HTML}{A6CE39}
\DeclareRobustCommand{\orcidicon}{
	\begin{tikzpicture}
	\draw[lime, fill=lime] (0,0) 
	circle [radius=0.16] 
	node[white] {{\fontfamily{qag}\selectfont \tiny ID}};
	\draw[white, fill=white] (-0.0625,0.095) 
	circle [radius=0.007];
	\end{tikzpicture}
	\hspace{-0.3mm}
}

\foreach \x in {A, ..., Z}{\expandafter\xdef\csname orcid\x\endcsname{\noexpand\href{https://orcid.org/\csname orcidauthor\x\endcsname}
			{\noexpand\orcidicon}}
}

\DeclareSIUnit\gauss{G}

\newcolumntype{L}[1]{>{\raggedright\let\newline\\\arraybackslash\hspace{0pt}}m{#1}}
\newcolumntype{C}[1]{>{\centering\let\newline\\\arraybackslash\hspace{0pt}}m{#1}}
\newcolumntype{R}[1]{>{\raggedleft\let\newline\\\arraybackslash\hspace{0pt}}m{#1}}

\newcommand{\BlueMotTransition}{${^1\mathrm{S}_0} - {^1\mathrm{P}_1}$ }

\newcommand{\PRLchangetext}[1]{\textcolor{black}{#1}} 
\newcommand{\AddAppendix}[1]{\textcolor{black}{#1}} 

\newcommand{\Templatetext}[1]{\textcolor{black}{#1}}
\newcommand{\texttense}[1]{\textcolor{black}{#1}}

\let\oldchi\chi
\renewcommand{\chi}{%
  \raisebox{0.44ex}{$\oldchi$}%
}

\begin{document}
\title{Narrow-line-mediated Sisyphus cooling in the $^{3}\mathrm{P}_{2}$ metastable state of strontium}
\author{Chun-Chia Chen (陳俊嘉)\,$\orcidA{}$,\textsuperscript{1,*\dag}, Ryoto Takeuchi\,$\orcidB{}$,\textsuperscript{2,3,*}, Shoichi Okaba\,$\orcidC{}$,\textsuperscript{1,2},  Hidetoshi Katori\,$\orcidD{}$,\textsuperscript{1,2,3\dag}}

\affiliation{\textsuperscript{1}Space-Time Engineering Research Team, RIKEN Center for Advanced Photonics, Wako, Saitama 351-0198, Japan}
\affiliation{\textsuperscript{2}Department of Applied Physics, Graduate School of Engineering, The University of Tokyo, Bunkyo-ku, Tokyo 113-8656, Japan}
\affiliation{\textsuperscript{3}Quantum Metrology Laboratory, RIKEN, Wako, Saitama 351-0198, Japan}

\date{\today}

\begin{abstract}
We demonstrate narrow-line-mediated Sisyphus cooling of magnetically trapped strontium (Sr) in the $5s5p\,^{3}\textrm{P}_{2}$ state. A \SI{641}{\nano\meter} standing-wave, blue-detuned from the $5s4d\,^{3}\textrm{D}_{3}$$\,\rightarrow$ $\,5p4d\,^{3}\textrm{F}_{4}$ transition, creates a dissipative optical lattice in the $^{3}\textrm{D}_{3}$ state. By combining Doppler cooling and Sisyphus cooling on the $5s5p\,^{3}\textrm{P}_{2}$$\,\rightarrow$ $5s4d\,^{3}\textrm{D}_{3}$ transition at $\SI{2.92}{\micro\meter}$, we observed efficient cooling of magnetically trapped atoms. 
By optically pumping the atoms to the $5s5p\,^{3}\textrm{P}_{0}$ state, we facilitate continuous outcoupling via a moving optical lattice with two fold improvement in atom number. 
Our scheme applies to next-generation quantum sensors using continuous ultracold atomic beams.
\end{abstract}

\begin{CJK*}{UTF8}{min}
\maketitle
\end{CJK*}

\footnote{\label{a}$^*$ These authors contributed equally}
\footnote{\label{b}$^{\dag}$ chenchunchia@gmail.com, katori@amo.t.u-tokyo.ac.jp}

A continuous ultracold 
atomic source is valuable for next-generation neutral atom quantum hardware~\cite{Chen2019Beam, Huntington2023_BufferGasBeam, Okaba2024SrBeam,cline2022continuous}. 
In quantum sensing applications, including optical clocks and atom interferometers, zero-dead-time operation improves their instability as $\tau^{-1}$ with an averaging time $\tau$~\cite{dick1987local, lodewyck2010frequency, Katori2021Srbeamclock}. This technique is at the forefront of quantum sensing research alongside other approaches such as spin squeezing~\cite{Pedrozo-Penafiel2020Squeezing, Eckner2023Squeezing, robinson2024direct}.
Motivated by similar performance scaling arguments toward better quantum hardware, neutral-atom-based quantum computing platforms are moving toward continuously reloading architecture to scale up atom arrays and realize the next generation large-scale logical qubits for fault-tolerant quantum computing applications~\cite{Pause2023CWReservoir, Gyger2024LargeArray, Norcia2024AtomComputing}.  

Preserving atomic coherence is crucial for next-generation continuous quantum devices. One approach leverages the ac-Stark shift to shield atoms from resonant light, which enables extending matter-wave coherence time in a continuous Bose-Einstein condensate~\cite{Chen2022_CWBEC} and protecting qubit coherence during mid-circuit quantum gate readout~\cite{Joanna2023_MidcircuitOmg, Norcia2023Midcircuit}. Another approach exploits long-lived metastable states.
They allow not only shelving of qubit coherence~\cite{Allcock2021, Joanna2023_MidcircuitOmg} but also continuous cold-atom generation through state-transfer techniques~\cite{Okaba2024SrBeam, Katori2001AIPLaserCoolingSr, 
Yang2007Ca3P2Loading, Chen2023CWBECreview,  Takeuchi2023CWSrBeam}. This approach particularly appeals to continuous clock operation with alkaline-earth(-like) elements~\cite{Hobson2020IRMOT, Akatsuka2021_3stageCooling,Katori2021Srbeamclock}.
Since the long-lived \({^3\mathrm{P}_{2}}\) metastable state~\cite{Yasuda2004Sr3P2} enables laser cooling on the \({^3\mathrm{P}_{2}}\rightarrow{^3\mathrm{D}_{3}}\) transition~\cite{Hemmerich2002Ca3P2, Hobson2020IRMOT, Akatsuka2021_3stageCooling,Takeuchi2023CWSrBeam}, which is independent of the \({^1\mathrm{S}_{0}}\rightarrow{^3\mathrm{P}_{0}}\) clock transition, simultaneous clock interrogation and atomic cooling are feasible.
Continuous generation of ultracold \({}^{88}\mathrm{Sr}\) was demonstrated
using two-stage laser cooling on the \({^1\mathrm{S}_{0}}\rightarrow{^1\mathrm{P}_{1}}\) and the narrow \({^3\mathrm{P}_{2}}\rightarrow{^3\mathrm{D}_{3}}\) transitions in a magnetic trap. 
Despite the low Doppler cooling limit for the narrow transition,
the Zeeman shift originating from the trap magnetic field prevented efficient laser cooling, requiring a long cooling time of a few seconds~\cite{Takeuchi2023CWSrBeam}.

Recently, narrow-line-mediated Sisyphus cooling has been demonstrated in various systems ranging from atomic beam deceleration~\cite{Chen2019SOLD} to cooling of trapped atoms to low temperatures~\cite{Cooper2018_SrTweezer, Chen2024ClockSisyphus}. 
In this Letter, we report a narrow-line-mediated Sisyphus cooling of magnetically trapped $^{88}$Sr atoms in the ${5s5p\,^3\mathrm{P}_{2}}$ state.
The Sisyphus mechanism is facilitated by a dissipative optical lattice in the $5s4d\,^{3}\textrm{D}_{3}$ state, which is formed by blue-detuned lasers from the $5s4d\,^{3}\textrm{D}_{3}$$\,\rightarrow$$5p4d\,^{3}\textrm{F}_{4}$ transition at \SI{641} {\nano\meter}.
By preferentially exciting atoms to the bottom of the optical lattice by the narrow-line $5s5p\,^{3}\textrm{P}_{2}$$\,\rightarrow$$5s4d\,^{3}\textrm{D}_{3}$ transition at \SI{2.92}{\micro\meter}, atoms are efficiently cooled through Sisyphus cycles. 
Moreover, by optically pumping the atoms to the $5s5p\,{^3\mathrm{P}_{0}}$ state, we demonstrate a continuous loading of atoms into a moving lattice with a doubled atom flux, compared to the previous work~\cite{Okaba2024SrBeam}.

\begin{figure}[t!]
\includegraphics[width=1\columnwidth]{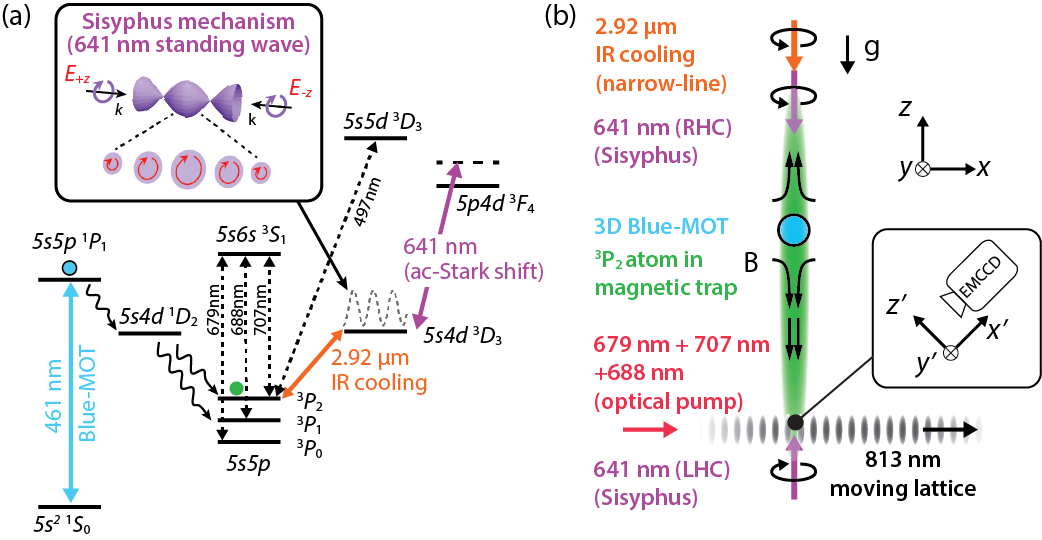}\caption{\label{fig:Setup_scheme} 
A narrow-line-mediated Sisyphus cooling on the $^{3}\textrm{P}_{2}\rightarrow{}^{3}\textrm{D}_{3}$ transition at 2.92\, \textmu m. (a) Relevant energy levels for $^{88}$Sr. The ‘blue’ (461\,nm), ‘IR’ (\SI{2.92}{\micro\meter}), and ‘Sisyphus’ (\SI{641}{\nano\meter}) transitions are used for continuous generation of ultracold $^{3}\mathrm{P}_{2}$ atoms. Transitions for optical pumping (679 nm, 688 nm, 707 nm) and imaging of $^{3}\textrm{P}_{2}$ atoms (497 nm) are also described.
(b) A pair of counter-propagating \SI{641}{\nano\meter} Sisyphus beams form a 1D standing wave in the $z$-axis. An IR laser at \SI{2.92}{\micro\meter} is introduced along the long axis of the elongated magnetic trap to cool the atoms in the $^{3}\mathrm{P}_{2}$ state and to load them into a moving lattice at \SI{813}{\nano\meter}. 
We take the center of the quadrupole magnetic field to be the origin of the coordinate $(x, y, z)$ with its $z$-axis opposite to gravity.
To address the atom image taken by a camera, we define a new coordinate $(x',y',z')$ which is rotated by $45^{\circ}$ around $y$ axis with its origin set at $z=-5\,{\rm mm}$.}
\end{figure}

\label{Sec:Method}
Our apparatus and the experimental procedures for magnetically trapping ultracold $^{88}$Sr have been described elsewhere~\cite{Takeuchi2023CWSrBeam}. A three-dimensional (3D) magneto-optical trap (MOT) cools and traps atoms on the \BlueMotTransition transition at \SI{461}{\nano\meter}, herein referred to as the blue-MOT.
As the $5s4d\,{}^1{\rm D}_2$ state lies below the $5s5p\,{}^1{\rm P}_1$ state~\cite{Hunter1986Sr1D2}, the atoms relaxed from the cooling transition are magnetically trapped in the long-lived $5s5p\,{^3\mathrm{P}_{2}}$ metastable state [see Fig.~\ref{fig:Setup_scheme}(a)]. 
Our magnetic trap has a prolate spheroid shape with magnetic field gradients of (10, 10, 0.7) $\mathrm{mT\,cm}^{-1}$ in the ($x,y,z$) directions, where we take the center of the quadrupole magnetic field to be the origin of the coordinate with the $z$-axis opposite to the gravity, as illustrated in Fig.~\ref{fig:Setup_scheme}(b). 

The small magnetic field gradient in the $z$-direction facilitates Doppler cooling on the ${^3\mathrm{P}_{2}}-{}^3\mathrm{D}_{3}$ transition with a narrow linewidth of \SI{57}{\kilo\hertz} at $\SI{2.92}{\micro\meter}$, which is referred to as IR cooling. A circularly polarized infrared (IR) cooling laser is applied in the $-z$ direction, providing a radiation pressure force $F_{\rm R}$. Combined with the magneto-gravitational force $F_{\rm BG}$~\cite{Takeuchi2023CWSrBeam}, the laser-cooled ${^3\mathrm{P}_{2}}$ atoms are trapped around the equilibrium position, satisfying $F_{\mathrm{BG}} + F_{\rm R} = 0$. A moving optical lattice at \SI{813}{\nano\meter} overlaps with the cooled atoms at $z_{\rm L}\sim -5\ \rm mm$, just below the center of the quadrupole magnetic field, facilitating the outcoupling of laser-cooled atoms.

\begin{figure}[t!]
\includegraphics[width=1\columnwidth]{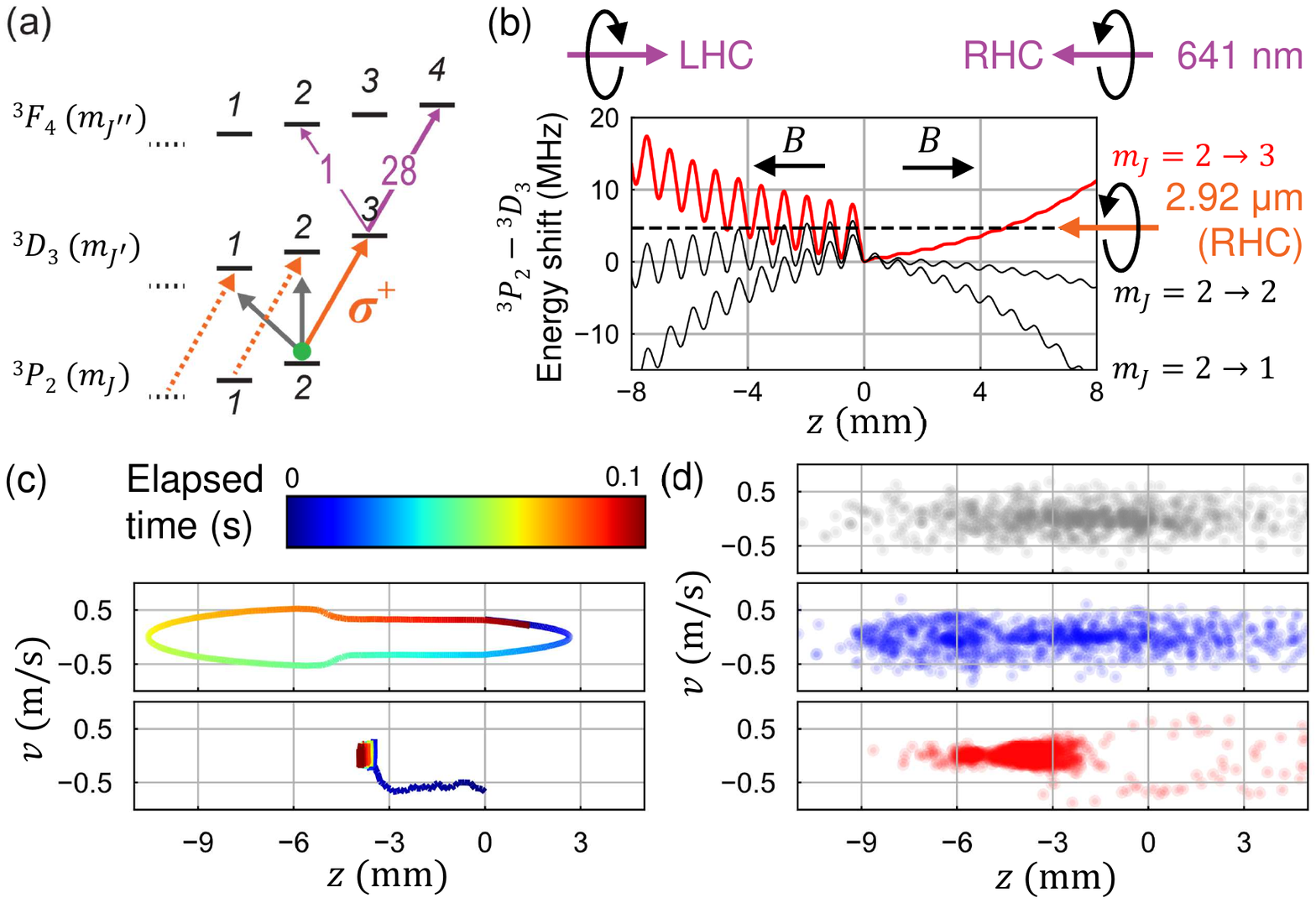}
\caption{\label{fig:simulation_main} 
Simulation results.
(a) The polarization of the IR laser is configured to address the $^{3}\mathrm{P}_{2}$$\,\rightarrow$\,$^{3}\textrm{D}_{3}$ $\sigma^{+}$  transition. For the circularly-polarized \SI{641}{\nano\meter} laser along the $z$-axis, it predominantly addresses $\sigma^+$ or $\sigma^-$ transition with distinct transition strengths as labeled next to the transition (purple line). (b) Energy shift for the $^3\mathrm{P}_{2}-{}^3\mathrm{D}_{3}$ transitions. Note that the Sisyphus potential period is not to scale. 
(c) Typical phase-space trajectories of an atom during Doppler cooling in the absence (upper) and presence (lower) of Sisyphus cooling. 
(d) Phase-space \PRLchangetext{distributions of} 1000 atoms. The upper panel shows the initial \PRLchangetext{distribution}. The middle (lower) panel shows \PRLchangetext{distribution} after applying Doppler cooling for 100~ms
in the absence (presence) of Sisyphus cooling.}
\end{figure}

Additionally, we introduce a pair of counterpropagating lasers at \SI{641}{\nano\meter} along the long axis of the elongated spherical-quadrupole magnetic field, ensuring overlap with the magnetically trapped atoms. These lasers, blue-detuned from the ${^3\mathrm{D}_3} - {^3\mathrm{F}_4}$ transition, introduce a spatially varying ac-Stark shift in the $^{3}\mathrm{D}_{3}$ state for Sisyphus cooling. They are focused to a $1/e^{2}$ diameter of approximately $\sim$ \SI{250}{\micro\meter} at $z_{\rm L}\sim -5\ \rm mm$ and
are circularly polarized as shown in Fig.~\ref{fig:Setup_scheme}(b). \PRLchangetext{For the magnetically trapped atoms located along the \SI{641}{\nano\meter} laser beam path with $z<0$, the local magnetic field is predominantly oriented along the $-z$ direction. This configuration defines the quantization axis such that the circularly polarized \SI{641}{\nano\meter} laser predominantly drives a $\sigma^{+}$ transition, resulting in a large ac-Stark shift in the ${^3\mathrm{D}_{3}}\, (m_J=3)$ state [see Fig.~\ref{fig:simulation_main}(a)]. In contrast, atoms at $z>0$ experience only a minimal shift due to the weaker coupling associated with the $\sigma^{-}$ transition [see Fig.~\ref{fig:simulation_main}(b)].} 

\begin{figure*}[t!]
\centering
\includegraphics[width=2\columnwidth]{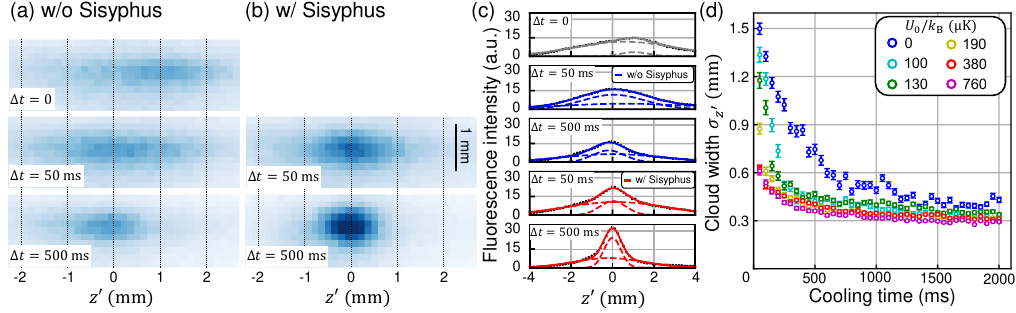}
\caption{\label{fig:3P2_SisyphusCoolinginMagneticTrap} 
Sisyphus cooling of magnetically trapped $^{88}$Sr in the $^{3}\mathrm{P}_{2}$ state. Fluorescence images of IR cooled atoms for $\Delta t=0$, 50~ms, and 500~ms (a) in the absence and (b) in the presence of Sisyphus cooling.  (c) Corresponding integrated spatial distribution of atoms along the $z'$-axis. The blue (red) lines correspond to IR cooling in the absence (presence) of Sisyphus cooling.
(d) Temporal evolution of the $1/\sqrt{e}$ cloud width $\sigma_{z'}$ along the $z'$-axis for various Sisyphus potential depths. The width is obtained from the Gaussian fit to the cold fraction of atoms when a bimodal distribution appears. 
}
\end{figure*}

\textit{Simulation.---}To understand the cooling mechanism, we conducted a one-dimensional (1D) simulation \AddAppendix{(see Appendix A)}~\cite{Sansonetti2010SrLines, Butcher2008RK, McClelland2008CrNarrowline}. We take into account the forces introduced by the magnetic trap, Sisyphus potential, and gravitational potential while the atoms interact with near-resonant light on the ${{^3\mathrm{P}_{2}} (m_J=2)} - {{^3\mathrm{D}_{3}} (m_J=3)}$ transition. We also include the recoil momentum transfer in the absorption and emission of IR photons.
Figure~\ref{fig:simulation_main}(c) shows the typical phase-space trajectories of an atom subjected to Doppler cooling in the absence (upper) and presence (lower) of Sisyphus cooling. 
In the Doppler cooling, the resonant conditions for the $^{3}\textrm{P}_{2}$$\,\rightarrow$$\,^{3}\textrm{D}_{3}$ transition are only satisfied near the position where the IR laser detuning matches the Zeeman shift~\cite{Takeuchi2023CWSrBeam}. 
Consequently, the cooling efficiency is limited as the atom experiences only a few cooling events every 0.1\,s that is given by the oscillation period of the atom in the magnetic trap in the $z$-direction.

In contrast, in the presence of Sisyphus potential, spatially modulated ac-Stark shift in the $^3\mathrm{D}_{3}$ state frequently brings atoms into resonance. 
Initially, an atom can be excited at the top of the Sisyphus potential [see Fig.~\ref{fig:simulation_main}(b) with $z\approx0$], leading to Sisyphus heating, as seen in the first few milliseconds of the trajectory in  Fig.~\ref{fig:simulation_main}(c). As the atom moves to regions where the IR laser excites the atom to the lower half of the Sisyphus potential~\cite{Chen2019SOLD}, a single-photon scattering event efficiently removes the kinetic energy, resulting in a rapid velocity decrease.  
We extended the simulation to 1,000 atoms as shown in Fig.~\ref{fig:simulation_main}(d), where we observed a significant increase in the atoms’ phase space density (PSD) after Sisyphus cooling of 100 ms as shown by the lower panel, sharply contrasting with the dynamics seen using Doppler cooling alone (see the middle panel).

\textit{Narrow-line-mediated Sisyphus cooling.---}We demonstrated narrow-line-mediated Sisyphus cooling of the magnetically trapped atoms. 
The ${^3\mathrm{P}_{2}}$ atoms \texttense{were} first loaded into the magnetic trap by operating the blue-MOT for 2 s.
After turning off the \SI{461}{\nano\meter} laser, we applied the IR cooling laser for a variable time.
Then we took fluorescence images of the atoms on the 
${^3\mathrm{P}_{2}} \rightarrow {5s5d\,^3\mathrm{D}_{3}}$ transition at \SI{497}{\nano\meter} using an electron multiplying CCD camera, where we frequency-modulated the laser by 80\,MHz to cover the Zeeman shift given by the magnetic trap. 
Due to optical access constraints, images are captured from a $45^{\circ}$ angle [Fig.~1(b)], defining a rotated coordinate system $(x',y',z')$ around the $y$-axis, with its origin set at $z=-5\,{\rm mm}$.

We captured the images after applying the IR cooling for $\Delta t=0$, 50~ms, and 500~ms as shown in Fig.~\ref{fig:3P2_SisyphusCoolinginMagneticTrap}(a).
The image with $\Delta t=0$ shows the initial distribution of atoms leaking out from the blue-MOT with the atomic temperature in the mK range.
We applied  IR cooling for 500\,ms with an intensity of \PRLchangetext{$I_{\mathrm{IR}} \sim 90\,I_{\mathrm{s}}^{\rm (IR)}$,
where $I_\mathrm{s}^{\rm (IR)}=\SI{0.3}{\micro\watt}/\rm cm^2$} is the saturation intensity of the IR transition. The atoms were gradually cooled and accumulated in the force balance position, which was shifted downward from the center of the magnetic trap.
As investigated in the previous studies \cite{Takeuchi2023CWSrBeam, Okaba2024SrBeam}, 
IR cooling required a few seconds to cool the atoms in the magnetic trap. 

Next, we applied Sisyphus cooling during the IR cooling.
A counter-propagating 641 nm laser formed the Sisyphus lattice with a potential depth of $U_{\rm 0}/k_{\rm B}=\SI{380}{\micro\kelvin}$.
Here we apply a lattice laser intensity of
\PRLchangetext{$I_{\mathrm{641}} \sim 2000\,I_{\rm s}^{(641)}$ with $I_{\rm s}^{(641)}=1.9\ \rm mW/cm^2$} the saturation intensity of the ${^3\mathrm{D}_{3}}-{^3\mathrm{F}_{4}}$ transition.
The laser was blue-detuned by $\Delta_{641}\approx2$ GHz from the resonance \AddAppendix{(see Appendix B)}. 
Under these conditions,  cold atoms rapidly accumulated in the force-balance trapping position, as seen in Fig.~\ref{fig:3P2_SisyphusCoolinginMagneticTrap}(b) at $\Delta t=50$~ms and $\Delta t=500$~ms, where the equilibrium position at $z'\approx0$~mm closely matches that of IR cooling shown in Fig.~\ref{fig:3P2_SisyphusCoolinginMagneticTrap}(a) after $\Delta t=500$~ms. 
Corresponding spatial distributions along the $z'$-axis are shown in Fig.~\ref{fig:3P2_SisyphusCoolinginMagneticTrap}(c).
It is noteworthy that the Sisyphus-cooled atoms developed a bimodal distribution with a distinct colder fraction after 50\,ms of cooling. 
\PRLchangetext{The hotter atoms remain as a broader distribution, which will be cooled in a multiple of oscillation periods along the $z$-axis of a few hundreds of ms.}

We studied the time evolution of the in-situ atomic cloud width inside the magnetic trap for different Sisyphus potential depth, as shown in Fig.~\ref{fig:3P2_SisyphusCoolinginMagneticTrap}(d).
The rapid reduction in cloud width highlights the efficiency of Sisyphus cooling.
Note that assuming the Gaussian distribution of an atomic cloud width of $\sigma_x$ and $\sigma_z$ along the $x$ and $z$ axes, the imaged width will be given by $\sigma_{z'}=\sqrt{(\sigma_x^2+\sigma_z^2)/2}$.
Thus the measurement resolution of $\sigma_{z'}$ is given by the $x$-width which is estimated to $\sigma_x\approx 0.3$~mm by taking the axial symmetry of the atomic cloud $\sigma_x\approx \sigma_y$ [see Fig.~3(a) and (b)].
The observed short cooling time of less than 0.1\,s agrees well with the simulations, as shown in  Fig.~\ref{fig:simulation_main}(d).
Such rapid cooling is beneficial for the efficient outcoupling of atoms into the moving lattice, as it prevents atoms from being lost by background-gas collision and light-assisted collisions~\cite{Akatsuka2021_3stageCooling} in the IR cooling time as long as a few seconds.
We conducted time-of-flight (TOF) measurements in the magnetic trap to determine the temperature of atoms along the $z$ axis after a cooling time of $1$ second. We estimated the temperature for the Sisyphus-cooled atoms to be $\sim$ 26\,\textmu K \AddAppendix{(see Appendix C)}, which is comparable to that achieved by IR Doppler cooling~\cite{Takeuchi2023CWSrBeam}.

\textit{Loading to the magic-wavelength optical lattice.---}
We demonstrate that the Sisyphus cooling enhances the loading efficiency of ${^3\mathrm{P}_{0}}$ atoms into the moving optical lattice formed in a bowtie build-up cavity~\cite{Okaba2024SrBeam}, where two 813\,nm lasers with a frequency difference of 
10\,kHz are coupled to provide the lattice velocity of 4 mm/s.
The moving lattice overlaps with the magnetic trap with a beam radius of {\SI{169}{\micro\meter}} and a trap depth of \SI{100}{\micro\kelvin} with the peak intensity of $\sim 160\, {\rm kW/cm}^2$.
Atoms in the $5p4d\,^{3}\textrm{F}_{4}$ state may be ionized by the lattice lasers at 813 nm, since the ionization threshold wavelength is estimated to be \SI{832}{\nano\meter}. 
To reduce the two-step ionization rate \(\bigl(\propto I_{641}/\Delta_{641}^2\bigr)\) via the $5s4d$ \(^3\mathrm{F}_4\) state while maintaining sufficient Sisyphus depth \(\bigl(\propto I_{641}/\Delta_{641}\bigr)\) for cooling, one may increase the detuning \(\Delta_{641}\) and the intensity \(I_{641}\) correspondingly.
Largely constrained by the available 641\,nm Sisyphus laser power, we blue-detuned the laser by $\Delta_{641}=5.6$\,GHz \AddAppendix{(see Table~\ref{tab:experiment_params} in Appendix D)}, which provided a Sisyphus trap depth of $U_{\rm 0}/k_{\rm B}\sim \SI{230}{\micro\kelvin}.$

\begin{figure}[t!]
\includegraphics[width=1\columnwidth]{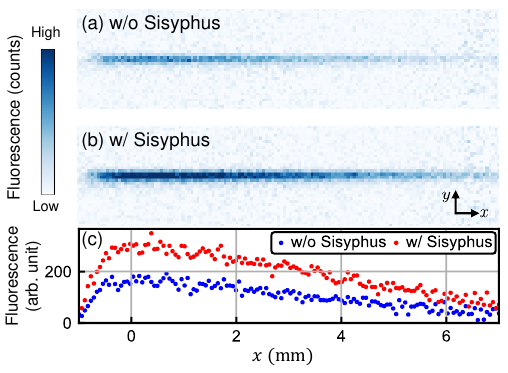}
\caption{\label{fig3:Lattice_loading} 
Continuous loading of ultracold $^{88}\mathrm{Sr}$ into the moving lattice at 813 nm. 
Fluorescence image of atoms in the moving lattice around the loading point $(x,y,z) = (0,0,-5\rm\,mm)$ in the absence (a) and presence (b) of Sisyphus cooling. 
(c) Integrated spatial distribution of atoms loaded into the moving-lattice.} 
\end{figure}

We conducted the continuous loading by applying the cooling lasers (461 nm, $\SI{2.92}{\micro\meter}$), the Sisyphus lattice (641~nm), the moving lattice (813~nm),
 and optical pumping lasers (688 nm and 707 nm) to populate the ${^3\mathrm{P}_{0}}$ state by optically pumping the atoms in the ${^3\mathrm{P}_{2}}$ state. 
To observe an in situ fluorescence image of atoms in the moving lattice with a 461 nm laser, 
we flashed two pumping lasers (679 nm and 707 nm), which were superimposed on the moving optical lattice, to pump the atoms from the ${^3\mathrm{P}_{0}}$ state to the ${^1\mathrm{S}_{0}}$ ground state.
As shown in Fig.~\ref{fig3:Lattice_loading}, an application of Sisyphus cooling increases the atom-loading efficiency by 88(10)\% by avoiding the collision loss of atoms occurring in the seconds-long IR cooling time.  The decrease of fluorescence intensity as \(x(=vt)\), with \(v = 4\,\mathrm{mm/s}\) the lattice velocity and \(t\) the elapsed time, as shown in Fig.~4(c), indicates the atom lifetime $\tau\sim 1.2$~s in the moving lattice, which is attributed to the background-gas-collision loss at a vacuum pressure of about $8 \times 10^{-8}$ Pa~\cite{Takeuchi2023CWSrBeam} and to the parametric heating loss of atoms from the lattice.

In summary, we have demonstrated narrow-line-mediated Sisyphus cooling of magnetically trapped $^{88}$Sr atoms in the $^3\mathrm{P}_{2}$ state using a dissipative optical lattice created in the ${5s4d\,^3\mathrm{D}_{3}}$ state by the standing-wave laser at 641\,nm.
We show that Sisyphus cooling reduces the cooling time by an order of magnitude. Consequently, the number of atoms outcoupled to the moving lattice in the $^{3}\textrm{P}_{0}$ state was nearly doubled.
Our work will be extended to a 3D Sisyphus cooling to further improve the outcoupling efficiency.
Our work paves the way for creating ultracold atomic sources suited for 
continuous optical clocks~\cite{Katori2021Srbeamclock} and continuous-wave superradiant lasers~\cite{chen2009active, meiser2009prospects, Norcia2016superradiance, kristensen2023subnatural}.

\section*{ACKNOWLEDGMENTS}

We thank M. Takamoto and S. Tsuji for experimental support and H. Chiba for assistance in experiments. We thank \Templatetext{S. Bennetts} for careful reading of the manuscript and providing useful comments. This work received support from the Japan Science and Technology Agency (JST) Mirai Program Grant No. JPMJMI18A1.




\section*{DATA AVALABILITY}
The data that support the findings of this article are openly available~\cite{dataavailability}.

\section*{APPENDIX A: SIMULATION}
We numerically simulated a one-dimensional cooling dynamics along the $z$ axis using the Runge-Kutta method~\cite{Butcher2008RK}.
The simulation accounts for the forces at each time step due to the gradients of the magnetic trap, Sisyphus lattice, and gravitational potential. We take into account the velocity changes given by the \SI{2.92}{\micro\meter} photons in the transitions between $\ket{0} = \ket{{^3\mathrm{P}_{2}},\ m_J=2}$ and $\ket{1} = \ket{{^3\mathrm{D}_{3}},\ m_J=3}$.

At each time step $i$, we use a vector $q_i=(z_i,v_i,s_i)$ consisting of position, velocity, and electronic state to represent an atom's status. From $q_{i}$, we calculate $q_{i+1}$ as follows:

1. $z_{i+1},v_{i+1}$ are updated by the Runge-Kutta method, where the acceleration $a(z,v,s)$ is calculated as
\begin{equation}
    a(z,v,s) = \left\{
            \begin{array}{ll}
                -\frac{3\mu_{\rm B}}{m}\dv{|B(z)|}{z} - g & (s = \ket{0})\\
                -\frac{4\mu_{\rm B}}{m}\dv{|B(z)|}{z} - \frac{1}{m}\dv{U_{\rm S}(z)}{z} - g & (s = \ket{1})
            \end{array}
            \right.
\end{equation}
Here, $\mu_{\rm B}$ is the Bohr magneton, $B(z)$ is the magnetic flux density at position $z$, $m$ is the mass of $^{88}\rm Sr$, $g$ is the gravitational acceleration, and $U_{\rm S}(z)=U_0\cos(2\pi z/\lambda_{641})^2$ is the Sisyphus lattice potential for the ${^3\mathrm{D}_{3}}( m_J=3)$ state.

2. The probability of photon absorption or emission $p(z,v,s)$ during a time interval $\Delta t$ is calculated as:
\begin{equation}
\begin{aligned}
    p(z,v,s) &= \left\{
            \begin{array}{ll}
               1 - \exp[-R(z,v)\Delta t ]& (s = \ket{0})\\
               1 - \exp[-\Gamma_{\rm IR}\Delta t ] & (s = \ket{1})
            \end{array}
            \right.\\
\end{aligned}
\end{equation}
where
\PRLchangetext{
\begin{equation}
\begin{aligned}
    R(z,v) & =\frac{\Gamma_{\rm IR}}{2} \frac{I/I_{\rm s}^{\rm (IR)}}{1+I/I_{\rm s}^{\rm (IR)}+[2\Delta(z,v)/\Gamma_{\rm IR}]^2}\\
    \Delta(z,v) &= -\delta \omega - k_{\rm IR}v + \mu_{\rm B}|B(z)|/\hbar + U_{\rm S}(z)/\hbar
\end{aligned}
\end{equation}
}
$\delta \omega$ is the detuning of the IR laser. \PRLchangetext{We choose $\Delta t=\SI{0.1}{\micro\second}$, which is an order of magnitude shorter than both the ${^3\mathrm{D}_{3}}$ lifetime ($ \approx \SI{2.8}{\micro\second}$) and the oscillation period of atoms in the Sisyphus lattice ($\approx \SI{2.5}{\micro\second}$).}

3. When the absorption or emission happens, $v_{i+1},s_{i+1}$ are updated as follows:
 \begin{equation}
 \begin{aligned}
    v_{i+1} &= \left\{
            \begin{array}{ll}
                v_{i+1} - \hbar k_{\rm IR}/m & (s_i = \ket{0})\\
                v_{i+1} + u\hbar k_{\rm IR}/m & (s_i = \ket{1})
            \end{array}
            \right.\\
    s_{i+1} &=  \left\{
            \begin{array}{ll}
                \ket{1} & (s_i = \ket{0})\\
                \ket{0} & (s_i = \ket{1})
            \end{array}
            \right.
\end{aligned}
\end{equation}
$u$ is a random value of either +1 or -1.

\begin{figure}
\centering
\includegraphics[width=0.98\columnwidth]{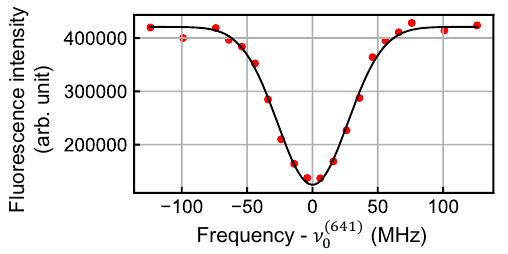}
\caption{\label{fig:641nmResonance} Spectra of the $5s4d\,^{3}\textrm{D}_{3}$$\,\rightarrow$$\,5p4d\,^{3}\textrm{F}_{4}$ transition measured by atom loss spectroscopy in a magnetic trap.  }
\end{figure}

\begin{figure}
\includegraphics[width=0.98\columnwidth]{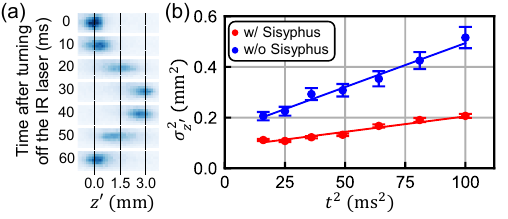}
\caption{\label{fig:osc} Oscillation of the atomic cloud in the magnetic trap. (a) Time evolution of Sisyphus-cooled atomic cloud oscillating in a magnetic trap after release. Fluorescence images are taken during the first oscillation period.
(b) The time-dependent width $\sigma_{z'}$ of atoms is plotted for the initial 10 ms of the oscillation. \color{black} We estimate the atomic temperatures $\sim$ 74\,\textmu K for the Doppler cooling, and $\sim$ 26\,\textmu K by applying the Sisyphus cooling.
}
\end{figure}

\section*{APPENDIX B: LOSS SPECTROSCOPY}
We conducted spectroscopy on the $5s4d\,^{3}\textrm{D}_{3}$$\,\rightarrow$$\,5p4d\,^{3}\textrm{F}_{4}$ transition at \SI{641}{\nano\meter} by simultaneously irradiating the ${^3\mathrm{P}_{2}}$($m_J$=2) atoms stored in a magnetic trap with the IR laser and \SI{641}{\nano\meter} laser. 
When the 641\,nm spectroscopy laser is tuned near resonance~\cite{Sansonetti2010SrLines}, atoms can scatter multiple 641\,nm photons that optically pump the atoms into the high-field-seeking ${^3\mathrm{P}_{2}}(m_J<0)$ states, resulting in the atom loss from the magnetic trap.  
We varied the frequency of the \SI{641}{\nano\meter} laser in consecutive measurements and characterized the number of atoms remaining in the magnetic trap as shown in Fig.~\ref{fig:641nmResonance}. We thus determined the resonance frequency $\nu_0^{(641)}$, which is used as a reference to determine the Sisyphus lattice detunings.

Initially, we began searching for the transition using the full power ($\sim$1\,mW) of the 641\,nm laser and gradually reduced the power to moderate the line broadening.
The laser frequencies were measured by a wavemeter that was referenced to the Sr $^{1}\textrm{S}_{0}$$\,\rightarrow$$^{3}\textrm{P}_{0}$ clock transition.

\section*{APPENDIX C: TIME-OF-FLIGHT IN THE MAGNETIC TRAP}
We conducted time-of-flight (TOF) measurements in the magnetic trap to determine the temperature of atoms along the $z$-axis after a cooling time of $1$ second. 
After turning off the IR laser, the atoms experienced a restoring force from the magneto-gravitational potential and began accelerating along the $z$-axis toward the center of the magnetic trap~\cite{McClelland2008CrNarrowline} while remaining confined along the other axes, as shown in Fig.~\ref{fig:osc}(a). We measured the oscillation period of the atomic cloud along the $z$-axis to be $T\approx 60$ ms.
For shorter flight times $t$, where the influence of the acceleration $a_{\rm BG}$ of atoms by the magneto-gravitational potential is smaller than the expansion of atoms with an initial velocity $v_0$, i.e., $\frac{1}{2}a_{\rm mg}t^2\ll v_{0} t$, the width of the atomic cloud indicates the atomic temperature. We note this temperature characterization may only serve as an underestimation, especially for longer $t$ and lower atomic temperatures.

We measured the $1/\sqrt{e}$ cloud width $\sigma_{z'}$ along the $z'$-axis by taking the narrower width of a double-Gaussian fit to the bimodal atomic distribution for $t<$10 ms after turning off the IR laser. 
In the presence of Sisyphus cooling, the initial width of the atomic cloud is smaller, and the width expands more slowly than that in Doppler cooling, indicating a lower temperature. 
We fit the temporal change of the squared cloud width $\sigma_{z'}^2$ using $\sigma_{z'}^2(t^2)=\sigma_{z'}^2(0)+k_{\rm B}T_{z}t^2/(2m)$ where $m$ is the mass of strontium, to estimate the temperature $T_z$ along the $z$-axis.
This time dependence arises from the fact that the width $\sigma_{z}$ along $z$-axis
expands as $\sigma_{z}^2(t^2)=\sigma_{z}^2(0)+k_{\rm B}T_{z}t^2/m$, while the width $\sigma_x$ along the $x$-axis remains constant, together with the relation $\sigma_{z'}^2=(\sigma_{x}^2+\sigma_{z}^2)/2$.
The estimated temperature for the Sisyphus-cooled atoms is $\sim$ 26\,\textmu K. In comparison, Doppler-cooled atoms exhibit a temperature of $\sim$ 74\,\textmu K, see Fig.~\ref{fig:osc}(b). This higher temperature for the Doppler-cooled sample reflects the slower cooling rate and the inability to reach a steady state within the 1-second cooling duration\cite{Takeuchi2023CWSrBeam}. 

\section*{APPENDIX D: TYPICAL PARAMETERS USED IN SISYPHUS COOLING}

\begin{table}[H]
\centering
\renewcommand{\arraystretch}{2.2}
\setlength{\tabcolsep}{5pt}
\begin{tabular}{lcccc}
Experiment & \makecell{$\Delta_{641}$\\{(GHz)}} &\PRLchangetext{$I_{641}/I_{\rm s}^{(641)}$}&\makecell{$R_{641}$\\$\rm (s^{-1})$} & \makecell{$U_0/k_{\rm B}$ \\$(\SI{}{\micro\kelvin})$}\\
\hline
\hline
\makecell{{Sisyphus}\\{cooling}} & 2.0  & $2.1\times10^3$ & $9.3\times10^4$ & 380\\
\makecell{{Lattice}\\{loading}} & 5.6 & $3.7\times10^3$ & $2.0\times10^4$ & 230\\
\hline
\end{tabular}
\caption{Typical parameters of the 641 nm laser used in our experiments. $\Delta_{641}$ is the detuning from the $5s4d\,^{3}\textrm{D}_{3}$$\,\rightarrow$$\,5p4d\,^{3}\textrm{F}_{4}$ resonance. The typical Sisyphus laser intensity ($I_{641}$) is shown for each counter-propagating beam. The scattering rate $R_{641}$ and the maximum Sisyphus lattice depth $U_0$ are calculated considering the interference enhancement between the counter-propagating beams.}
\label{tab:experiment_params}
\end{table}


\begin{thebibliography}{37}%
\makeatletter
\providecommand \@ifxundefined [1]{%
 \@ifx{#1\undefined}
}%
\providecommand \@ifnum [1]{%
 \ifnum #1\expandafter \@firstoftwo
 \else \expandafter \@secondoftwo
 \fi
}%
\providecommand \@ifx [1]{%
 \ifx #1\expandafter \@firstoftwo
 \else \expandafter \@secondoftwo
 \fi
}%
\providecommand \natexlab [1]{#1}%
\providecommand \enquote  [1]{``#1''}%
\providecommand \bibnamefont  [1]{#1}%
\providecommand \bibfnamefont [1]{#1}%
\providecommand \citenamefont [1]{#1}%
\providecommand \href@noop [0]{\@secondoftwo}%
\providecommand \href [0]{\begingroup \@sanitize@url \@href}%
\providecommand \@href[1]{\@@startlink{#1}\@@href}%
\providecommand \@@href[1]{\endgroup#1\@@endlink}%
\providecommand \@sanitize@url [0]{\catcode `\\12\catcode `\$12\catcode `\&12\catcode `\#12\catcode `\^12\catcode `\_12\catcode `\%12\relax}%
\providecommand \@@startlink[1]{}%
\providecommand \@@endlink[0]{}%
\providecommand \url  [0]{\begingroup\@sanitize@url \@url }%
\providecommand \@url [1]{\endgroup\@href {#1}{\urlprefix }}%
\providecommand \urlprefix  [0]{URL }%
\providecommand \Eprint [0]{\href }%
\providecommand \doibase [0]{http://dx.doi.org/}%
\providecommand \selectlanguage [0]{\@gobble}%
\providecommand \bibinfo  [0]{\@secondoftwo}%
\providecommand \bibfield  [0]{\@secondoftwo}%
\providecommand \translation [1]{[#1]}%
\providecommand \BibitemOpen [0]{}%
\providecommand \bibitemStop [0]{}%
\providecommand \bibitemNoStop [0]{.\EOS\space}%
\providecommand \EOS [0]{\spacefactor3000\relax}%
\providecommand \BibitemShut  [1]{\csname bibitem#1\endcsname}%
\let\auto@bib@innerbib\@empty

\bibitem [{\citenamefont {Chen}\ \emph {et~al.}(2019{\natexlab{a}})\citenamefont {Chen}, \citenamefont {Bennetts}, \citenamefont {Escudero}, \citenamefont {Pasquiou},\ and\ \citenamefont {Schreck}}]{Chen2019Beam}%
  \BibitemOpen
  \bibfield  {author} {\bibinfo {author} {\bibfnamefont {Chun-Chia}\ \bibnamefont {Chen}}, \bibinfo {author} {\bibfnamefont {Shayne}\ \bibnamefont {Bennetts}}, \bibinfo {author} {\bibfnamefont {Rodrigo~Gonz\'alez}\ \bibnamefont {Escudero}}, \bibinfo {author} {\bibfnamefont {Benjamin}\ \bibnamefont {Pasquiou}}, \ and\ \bibinfo {author} {\bibfnamefont {Florian}\ \bibnamefont {Schreck}},\ }\bibfield  {title} {\enquote {\bibinfo {title} {Continuous guided strontium beam with high phase-space density},}\ }\href {\doibase 10.1103/PhysRevApplied.12.044014} {\bibfield  {journal} {\bibinfo  {journal} {Phys. Rev. Appl.}\ }\textbf {\bibinfo {volume} {12}},\ \bibinfo {pages} {044014} (\bibinfo {year} {2019}{\natexlab{a}})}\BibitemShut {NoStop}%
\bibitem [{\citenamefont {Huntington}\ \emph {et~al.}(2023)\citenamefont {Huntington}, \citenamefont {Glick}, \citenamefont {Borysow},\ and\ \citenamefont {Heinzen}}]{Huntington2023_BufferGasBeam}%
  \BibitemOpen
  \bibfield  {author} {\bibinfo {author} {\bibfnamefont {William}\ \bibnamefont {Huntington}}, \bibinfo {author} {\bibfnamefont {Jeremy}\ \bibnamefont {Glick}}, \bibinfo {author} {\bibfnamefont {Michael}\ \bibnamefont {Borysow}}, \ and\ \bibinfo {author} {\bibfnamefont {Daniel~J.}\ \bibnamefont {Heinzen}},\ }\bibfield  {title} {\enquote {\bibinfo {title} {Intense continuous cold-atom source},}\ }\href {\doibase 10.1103/PhysRevA.107.013302} {\bibfield  {journal} {\bibinfo  {journal} {Phys. Rev. A}\ }\textbf {\bibinfo {volume} {107}},\ \bibinfo {pages} {013302} (\bibinfo {year} {2023})}\BibitemShut {NoStop}%
\bibitem [{\citenamefont {Okaba}\ \emph {et~al.}(2024)\citenamefont {Okaba}, \citenamefont {Takeuchi}, \citenamefont {Tsuji},\ and\ \citenamefont {Katori}}]{Okaba2024SrBeam}%
  \BibitemOpen
  \bibfield  {author} {\bibinfo {author} {\bibfnamefont {Shoichi}\ \bibnamefont {Okaba}}, \bibinfo {author} {\bibfnamefont {Ryoto}\ \bibnamefont {Takeuchi}}, \bibinfo {author} {\bibfnamefont {Shigenori}\ \bibnamefont {Tsuji}}, \ and\ \bibinfo {author} {\bibfnamefont {Hidetoshi}\ \bibnamefont {Katori}},\ }\bibfield  {title} {\enquote {\bibinfo {title} {Continuous generation of an ultracold atomic beam using crossed moving optical lattices},}\ }\href {\doibase 10.1103/PhysRevApplied.21.034006} {\bibfield  {journal} {\bibinfo  {journal} {Phys. Rev. Appl.}\ }\textbf {\bibinfo {volume} {21}},\ \bibinfo {pages} {034006} (\bibinfo {year} {2024})}\BibitemShut {NoStop}%
\bibitem [{\citenamefont {Cline}\ \emph {et~al.}(2025)\citenamefont {Cline}, \citenamefont {Sch\"afer}, \citenamefont {Niu}, \citenamefont {Young}, \citenamefont {Yoon},\ and\ \citenamefont {Thompson}}]{cline2022continuous}%
  \BibitemOpen
  \bibfield  {author} {\bibinfo {author} {\bibfnamefont {Julia R.~K.}\ \bibnamefont {Cline}}, \bibinfo {author} {\bibfnamefont {Vera~M.}\ \bibnamefont {Sch\"afer}}, \bibinfo {author} {\bibfnamefont {Zhijing}\ \bibnamefont {Niu}}, \bibinfo {author} {\bibfnamefont {Dylan~J.}\ \bibnamefont {Young}}, \bibinfo {author} {\bibfnamefont {Tai~Hyun}\ \bibnamefont {Yoon}}, \ and\ \bibinfo {author} {\bibfnamefont {James~K.}\ \bibnamefont {Thompson}},\ }\bibfield  {title} {\enquote {\bibinfo {title} {Continuous collective strong coupling of strontium atoms to a high finesse ring cavity},}\ }\href {\doibase 10.1103/PhysRevLett.134.013403} {\bibfield  {journal} {\bibinfo  {journal} {Phys. Rev. Lett.}\ }\textbf {\bibinfo {volume} {134}},\ \bibinfo {pages} {013403} (\bibinfo {year} {2025})}\BibitemShut {NoStop}%
\bibitem [{\citenamefont {Dick}\ \emph {et~al.}(1987)\citenamefont {Dick}, \citenamefont {Prestage}, \citenamefont {Greenhall},\ and\ \citenamefont {Maleki}}]{dick1987local}%
  \BibitemOpen
  \bibfield  {author} {\bibinfo {author} {\bibfnamefont {G.~J.}\ \bibnamefont {Dick}}, \bibinfo {author} {\bibfnamefont {J.~D.}\ \bibnamefont {Prestage}}, \bibinfo {author} {\bibfnamefont {C.~A.}\ \bibnamefont {Greenhall}}, \ and\ \bibinfo {author} {\bibfnamefont {L.}~\bibnamefont {Maleki}},\ }\bibfield  {title} {\enquote {\bibinfo {title} {Local oscillator induced degradation of medium-term stability in passive atomic frequency standards},}\ }\href@noop {} {\bibfield  {journal} {\bibinfo  {journal} {{\it Proceedings of the 19th Annual Precise Time and Time Interval Systems and Applications Meeting}}\ ,\ \bibinfo {pages} {133--147}} (\bibinfo {year} {1987})}\BibitemShut {NoStop}%
\bibitem [{\citenamefont {Lodewyck}\ \emph {et~al.}(2010)\citenamefont {Lodewyck}, \citenamefont {Westergaard}, \citenamefont {Lecallier}, \citenamefont {Lorini},\ and\ \citenamefont {Lemonde}}]{lodewyck2010frequency}%
  \BibitemOpen
  \bibfield  {author} {\bibinfo {author} {\bibfnamefont {J{\'e}r{\^o}me}\ \bibnamefont {Lodewyck}}, \bibinfo {author} {\bibfnamefont {Philip~G}\ \bibnamefont {Westergaard}}, \bibinfo {author} {\bibfnamefont {Arnaud}\ \bibnamefont {Lecallier}}, \bibinfo {author} {\bibfnamefont {Luca}\ \bibnamefont {Lorini}}, \ and\ \bibinfo {author} {\bibfnamefont {Pierre}\ \bibnamefont {Lemonde}},\ }\bibfield  {title} {\enquote {\bibinfo {title} {Frequency stability of optical lattice clocks},}\ }\href {\doibase 10.1088/1367-2630/12/6/065026} {\bibfield  {journal} {\bibinfo  {journal} {New J. Phys.}\ }\textbf {\bibinfo {volume} {12}},\ \bibinfo {pages} {065026} (\bibinfo {year} {2010})}\BibitemShut {NoStop}%
\bibitem [{\citenamefont {Katori}(2021)}]{Katori2021Srbeamclock}%
  \BibitemOpen
  \bibfield  {author} {\bibinfo {author} {\bibfnamefont {Hidetoshi}\ \bibnamefont {Katori}},\ }\bibfield  {title} {\enquote {\bibinfo {title} {Longitudinal {Ramsey} spectroscopy of atoms for continuous operation of optical clocks},}\ }\href {\doibase 10.35848/1882-0786/ac0e16} {\bibfield  {journal} {\bibinfo  {journal} {Appl. Phys. Express}\ }\textbf {\bibinfo {volume} {14}},\ \bibinfo {pages} {072006} (\bibinfo {year} {2021})}\BibitemShut {NoStop}%
\bibitem [{\citenamefont {Pedrozo-Peñafiel}\ \emph {et~al.}(2020)\citenamefont {Pedrozo-Peñafiel}, \citenamefont {Colombo}, \citenamefont {Shu}, \citenamefont {Adiyatullin}, \citenamefont {Li}, \citenamefont {Mendez}, \citenamefont {Braverman}, \citenamefont {Kawasaki}, \citenamefont {Akamatsu}, \citenamefont {Xiao},\ and\ \citenamefont {Vuletić}}]{Pedrozo-Penafiel2020Squeezing}%
  \BibitemOpen
  \bibfield  {author} {\bibinfo {author} {\bibfnamefont {Edwin}\ \bibnamefont {Pedrozo-Peñafiel}}, \bibinfo {author} {\bibfnamefont {Simone}\ \bibnamefont {Colombo}}, \bibinfo {author} {\bibfnamefont {Chi}\ \bibnamefont {Shu}}, \bibinfo {author} {\bibfnamefont {Albert~F}\ \bibnamefont {Adiyatullin}}, \bibinfo {author} {\bibfnamefont {Zeyang}\ \bibnamefont {Li}}, \bibinfo {author} {\bibfnamefont {Enrique}\ \bibnamefont {Mendez}}, \bibinfo {author} {\bibfnamefont {Boris}\ \bibnamefont {Braverman}}, \bibinfo {author} {\bibfnamefont {Akio}\ \bibnamefont {Kawasaki}}, \bibinfo {author} {\bibfnamefont {Daisuke}\ \bibnamefont {Akamatsu}}, \bibinfo {author} {\bibfnamefont {Yanhong}\ \bibnamefont {Xiao}}, \ and\ \bibinfo {author} {\bibfnamefont {Vladan}\ \bibnamefont {Vuletić}},\ }\bibfield  {title} {\enquote {\bibinfo {title} {Entanglement on an optical atomic-clock transition},}\ }\href {\doibase 10.1038/s41586-020-3006-1} {\bibfield  {journal} {\bibinfo  {journal} {Nature}\ }\textbf {\bibinfo {volume} {588}},\
  \bibinfo {pages} {414--418} (\bibinfo {year} {2020})}\BibitemShut {NoStop}%
\bibitem [{\citenamefont {Eckner}\ \emph {et~al.}(2023)\citenamefont {Eckner}, \citenamefont {Darkwah~Oppong}, \citenamefont {Cao}, \citenamefont {Young}, \citenamefont {Milner}, \citenamefont {Robinson}, \citenamefont {Ye},\ and\ \citenamefont {Kaufman}}]{Eckner2023Squeezing}%
  \BibitemOpen
  \bibfield  {author} {\bibinfo {author} {\bibfnamefont {William~J.}\ \bibnamefont {Eckner}}, \bibinfo {author} {\bibfnamefont {Nelson}\ \bibnamefont {Darkwah~Oppong}}, \bibinfo {author} {\bibfnamefont {Alec}\ \bibnamefont {Cao}}, \bibinfo {author} {\bibfnamefont {Aaron~W.}\ \bibnamefont {Young}}, \bibinfo {author} {\bibfnamefont {William~R.}\ \bibnamefont {Milner}}, \bibinfo {author} {\bibfnamefont {John~M.}\ \bibnamefont {Robinson}}, \bibinfo {author} {\bibfnamefont {Jun}\ \bibnamefont {Ye}}, \ and\ \bibinfo {author} {\bibfnamefont {Adam~M.}\ \bibnamefont {Kaufman}},\ }\bibfield  {title} {\enquote {\bibinfo {title} {Realizing spin squeezing with rydberg interactions in an optical clock},}\ }\href {\doibase 10.1038/s41586-023-06360-6} {\bibfield  {journal} {\bibinfo  {journal} {Nature}\ }\textbf {\bibinfo {volume} {621}},\ \bibinfo {pages} {734--739} (\bibinfo {year} {2023})}\BibitemShut {NoStop}%
\bibitem [{\citenamefont {Robinson}\ \emph {et~al.}(2024)\citenamefont {Robinson}, \citenamefont {Miklos}, \citenamefont {Tso}, \citenamefont {Kennedy}, \citenamefont {Bothwell}, \citenamefont {Kedar}, \citenamefont {Thompson},\ and\ \citenamefont {Ye}}]{robinson2024direct}%
  \BibitemOpen
  \bibfield  {author} {\bibinfo {author} {\bibfnamefont {John~M.}\ \bibnamefont {Robinson}}, \bibinfo {author} {\bibfnamefont {Maya}\ \bibnamefont {Miklos}}, \bibinfo {author} {\bibfnamefont {Yee~Ming}\ \bibnamefont {Tso}}, \bibinfo {author} {\bibfnamefont {Colin~J.}\ \bibnamefont {Kennedy}}, \bibinfo {author} {\bibfnamefont {Tobias}\ \bibnamefont {Bothwell}}, \bibinfo {author} {\bibfnamefont {Dhruv}\ \bibnamefont {Kedar}}, \bibinfo {author} {\bibfnamefont {James~K.}\ \bibnamefont {Thompson}}, \ and\ \bibinfo {author} {\bibfnamefont {Jun}\ \bibnamefont {Ye}},\ }\bibfield  {title} {\enquote {\bibinfo {title} {Direct comparison of two spin-squeezed optical clock ensembles at the ${10}^{\ensuremath{-}17}$ level},}\ }\href {\doibase 10.1038/s41567-023-02310-1} {\bibfield  {journal} {\bibinfo  {journal} {Nat. Phys.}\ }\textbf {\bibinfo {volume} {20}},\ \bibinfo {pages} {208--213} (\bibinfo {year} {2024})}\BibitemShut {NoStop}%
\bibitem [{\citenamefont {Pause}\ \emph {et~al.}(2023)\citenamefont {Pause}, \citenamefont {Preuschoff}, \citenamefont {Sch\"affner}, \citenamefont {Schlosser},\ and\ \citenamefont {Birkl}}]{Pause2023CWReservoir}%
  \BibitemOpen
  \bibfield  {author} {\bibinfo {author} {\bibfnamefont {Lars}\ \bibnamefont {Pause}}, \bibinfo {author} {\bibfnamefont {Tilman}\ \bibnamefont {Preuschoff}}, \bibinfo {author} {\bibfnamefont {Dominik}\ \bibnamefont {Sch\"affner}}, \bibinfo {author} {\bibfnamefont {Malte}\ \bibnamefont {Schlosser}}, \ and\ \bibinfo {author} {\bibfnamefont {Gerhard}\ \bibnamefont {Birkl}},\ }\bibfield  {title} {\enquote {\bibinfo {title} {Reservoir-based deterministic loading of single-atom tweezer arrays},}\ }\href {\doibase 10.1103/PhysRevResearch.5.L032009} {\bibfield  {journal} {\bibinfo  {journal} {Phys. Rev. Res.}\ }\textbf {\bibinfo {volume} {5}},\ \bibinfo {pages} {L032009} (\bibinfo {year} {2023})}\BibitemShut {NoStop}%
\bibitem [{\citenamefont {Gyger}\ \emph {et~al.}(2024)\citenamefont {Gyger}, \citenamefont {Ammenwerth}, \citenamefont {Tao}, \citenamefont {Timme}, \citenamefont {Snigirev}, \citenamefont {Bloch},\ and\ \citenamefont {Zeiher}}]{Gyger2024LargeArray}%
  \BibitemOpen
  \bibfield  {author} {\bibinfo {author} {\bibfnamefont {Flavien}\ \bibnamefont {Gyger}}, \bibinfo {author} {\bibfnamefont {Maximilian}\ \bibnamefont {Ammenwerth}}, \bibinfo {author} {\bibfnamefont {Renhao}\ \bibnamefont {Tao}}, \bibinfo {author} {\bibfnamefont {Hendrik}\ \bibnamefont {Timme}}, \bibinfo {author} {\bibfnamefont {Stepan}\ \bibnamefont {Snigirev}}, \bibinfo {author} {\bibfnamefont {Immanuel}\ \bibnamefont {Bloch}}, \ and\ \bibinfo {author} {\bibfnamefont {Johannes}\ \bibnamefont {Zeiher}},\ }\bibfield  {title} {\enquote {\bibinfo {title} {Continuous operation of large-scale atom arrays in optical lattices},}\ }\href {\doibase 10.1103/PhysRevResearch.6.033104} {\bibfield  {journal} {\bibinfo  {journal} {Phys. Rev. Res.}\ }\textbf {\bibinfo {volume} {6}},\ \bibinfo {pages} {033104} (\bibinfo {year} {2024})}\BibitemShut {NoStop}%
\bibitem [{\citenamefont {Norcia}\ \emph {et~al.}(2024)\citenamefont {Norcia}, \citenamefont {Kim}, \citenamefont {Cairncross}, \citenamefont {Stone}, \citenamefont {Ryou}, \citenamefont {Jaffe}, \citenamefont {Brown}, \citenamefont {Barnes}, \citenamefont {Battaglino}, \citenamefont {Bohdanowicz}, \citenamefont {Brown}, \citenamefont {Cassella}, \citenamefont {Chen}, \citenamefont {Coxe}, \citenamefont {Crow}, \citenamefont {Epstein}, \citenamefont {Griger}, \citenamefont {Halperin}, \citenamefont {Hummel}, \citenamefont {Jones}, \citenamefont {Kindem}, \citenamefont {King}, \citenamefont {Kotru}, \citenamefont {Lauigan}, \citenamefont {Li}, \citenamefont {Lu}, \citenamefont {Megidish}, \citenamefont {Marjanovic}, \citenamefont {McDonald}, \citenamefont {Mittiga}, \citenamefont {Muniz}, \citenamefont {Narayanaswami}, \citenamefont {Nishiguchi}, \citenamefont {Paule}, \citenamefont {Pawlak}, \citenamefont {Peng}, \citenamefont {Pudenz}, \citenamefont {Rodr\'{\i}guez~P\'erez}, \citenamefont {Smull},
  \citenamefont {Stack}, \citenamefont {Urbanek}, \citenamefont {van~de Veerdonk}, \citenamefont {Vendeiro}, \citenamefont {Wadleigh}, \citenamefont {Wilkason}, \citenamefont {Wu}, \citenamefont {Xie}, \citenamefont {Zalys-Geller}, \citenamefont {Zhang},\ and\ \citenamefont {Bloom}}]{Norcia2024AtomComputing}%
  \BibitemOpen
  \bibfield  {author} {\bibinfo {author} {\bibfnamefont {M.~A.}\ \bibnamefont {Norcia}}, \bibinfo {author} {\bibfnamefont {H.}~\bibnamefont {Kim}}, \bibinfo {author} {\bibfnamefont {W.~B.}\ \bibnamefont {Cairncross}}, \bibinfo {author} {\bibfnamefont {M.}~\bibnamefont {Stone}}, \bibinfo {author} {\bibfnamefont {A.}~\bibnamefont {Ryou}}, \bibinfo {author} {\bibfnamefont {M.}~\bibnamefont {Jaffe}}, \bibinfo {author} {\bibfnamefont {M.~O.}\ \bibnamefont {Brown}}, \bibinfo {author} {\bibfnamefont {K.}~\bibnamefont {Barnes}}, \bibinfo {author} {\bibfnamefont {P.}~\bibnamefont {Battaglino}}, \bibinfo {author} {\bibfnamefont {T.~C.}\ \bibnamefont {Bohdanowicz}}, \bibinfo {author} {\bibfnamefont {A.}~\bibnamefont {Brown}}, \bibinfo {author} {\bibfnamefont {K.}~\bibnamefont {Cassella}}, \bibinfo {author} {\bibfnamefont {C.-A.}\ \bibnamefont {Chen}}, \bibinfo {author} {\bibfnamefont {R.}~\bibnamefont {Coxe}}, \bibinfo {author} {\bibfnamefont {D.}~\bibnamefont {Crow}}, \bibinfo {author} {\bibfnamefont {J.}~\bibnamefont
  {Epstein}}, \bibinfo {author} {\bibfnamefont {C.}~\bibnamefont {Griger}}, \bibinfo {author} {\bibfnamefont {E.}~\bibnamefont {Halperin}}, \bibinfo {author} {\bibfnamefont {F.}~\bibnamefont {Hummel}}, \bibinfo {author} {\bibfnamefont {A.~M.~W.}\ \bibnamefont {Jones}}, \bibinfo {author} {\bibfnamefont {J.~M.}\ \bibnamefont {Kindem}}, \bibinfo {author} {\bibfnamefont {J.}~\bibnamefont {King}}, \bibinfo {author} {\bibfnamefont {K.}~\bibnamefont {Kotru}}, \bibinfo {author} {\bibfnamefont {J.}~\bibnamefont {Lauigan}}, \bibinfo {author} {\bibfnamefont {M.}~\bibnamefont {Li}}, \bibinfo {author} {\bibfnamefont {M.}~\bibnamefont {Lu}}, \bibinfo {author} {\bibfnamefont {E.}~\bibnamefont {Megidish}}, \bibinfo {author} {\bibfnamefont {J.}~\bibnamefont {Marjanovic}}, \bibinfo {author} {\bibfnamefont {M.}~\bibnamefont {McDonald}}, \bibinfo {author} {\bibfnamefont {T.}~\bibnamefont {Mittiga}}, \bibinfo {author} {\bibfnamefont {J.~A.}\ \bibnamefont {Muniz}}, \bibinfo {author} {\bibfnamefont {S.}~\bibnamefont
  {Narayanaswami}}, \bibinfo {author} {\bibfnamefont {C.}~\bibnamefont {Nishiguchi}}, \bibinfo {author} {\bibfnamefont {T.}~\bibnamefont {Paule}}, \bibinfo {author} {\bibfnamefont {K.~A.}\ \bibnamefont {Pawlak}}, \bibinfo {author} {\bibfnamefont {L.~S.}\ \bibnamefont {Peng}}, \bibinfo {author} {\bibfnamefont {K.~L.}\ \bibnamefont {Pudenz}}, \bibinfo {author} {\bibfnamefont {D.}~\bibnamefont {Rodr\'{\i}guez~P\'erez}}, \bibinfo {author} {\bibfnamefont {A.}~\bibnamefont {Smull}}, \bibinfo {author} {\bibfnamefont {D.}~\bibnamefont {Stack}}, \bibinfo {author} {\bibfnamefont {M.}~\bibnamefont {Urbanek}}, \bibinfo {author} {\bibfnamefont {R.~J.~M.}\ \bibnamefont {van~de Veerdonk}}, \bibinfo {author} {\bibfnamefont {Z.}~\bibnamefont {Vendeiro}}, \bibinfo {author} {\bibfnamefont {L.}~\bibnamefont {Wadleigh}}, \bibinfo {author} {\bibfnamefont {T.}~\bibnamefont {Wilkason}}, \bibinfo {author} {\bibfnamefont {T.-Y.}\ \bibnamefont {Wu}}, \bibinfo {author} {\bibfnamefont {X.}~\bibnamefont {Xie}}, \bibinfo {author}
  {\bibfnamefont {E.}~\bibnamefont {Zalys-Geller}}, \bibinfo {author} {\bibfnamefont {X.}~\bibnamefont {Zhang}}, \ and\ \bibinfo {author} {\bibfnamefont {B.~J.}\ \bibnamefont {Bloom}},\ }\bibfield  {title} {\enquote {\bibinfo {title} {Iterative assembly of ${}^{171}$$\mathrm{Yb}$ atom arrays with cavity-enhanced optical lattices},}\ }\href {\doibase 10.1103/PRXQuantum.5.030316} {\bibfield  {journal} {\bibinfo  {journal} {PRX Quantum}\ }\textbf {\bibinfo {volume} {5}},\ \bibinfo {pages} {030316} (\bibinfo {year} {2024})}\BibitemShut {NoStop}%
\bibitem [{\citenamefont {Chen}\ \emph {et~al.}(2022)\citenamefont {Chen}, \citenamefont {Gonz{\'a}lez~Escudero}, \citenamefont {Min{\'a}{\v{r}}}, \citenamefont {Pasquiou}, \citenamefont {Bennetts},\ and\ \citenamefont {Schreck}}]{Chen2022_CWBEC}%
  \BibitemOpen
  \bibfield  {author} {\bibinfo {author} {\bibfnamefont {Chun-Chia}\ \bibnamefont {Chen}}, \bibinfo {author} {\bibfnamefont {Rodrigo}\ \bibnamefont {Gonz{\'a}lez~Escudero}}, \bibinfo {author} {\bibfnamefont {Ji{\v{r}}{\'i}}\ \bibnamefont {Min{\'a}{\v{r}}}}, \bibinfo {author} {\bibfnamefont {Benjamin}\ \bibnamefont {Pasquiou}}, \bibinfo {author} {\bibfnamefont {Shayne}\ \bibnamefont {Bennetts}}, \ and\ \bibinfo {author} {\bibfnamefont {Florian}\ \bibnamefont {Schreck}},\ }\bibfield  {title} {\enquote {\bibinfo {title} {Continuous {B}ose–{E}instein condensation},}\ }\href {\doibase 10.1038/s41586-022-04731-z} {\bibfield  {journal} {\bibinfo  {journal} {Nature}\ }\textbf {\bibinfo {volume} {606}},\ \bibinfo {pages} {683--687} (\bibinfo {year} {2022})}\BibitemShut {NoStop}%
\bibitem [{\citenamefont {Lis}\ \emph {et~al.}(2023)\citenamefont {Lis}, \citenamefont {Senoo}, \citenamefont {McGrew}, \citenamefont {R\"onchen}, \citenamefont {Jenkins},\ and\ \citenamefont {Kaufman}}]{Joanna2023_MidcircuitOmg}%
  \BibitemOpen
  \bibfield  {author} {\bibinfo {author} {\bibfnamefont {Joanna~W.}\ \bibnamefont {Lis}}, \bibinfo {author} {\bibfnamefont {Aruku}\ \bibnamefont {Senoo}}, \bibinfo {author} {\bibfnamefont {William~F.}\ \bibnamefont {McGrew}}, \bibinfo {author} {\bibfnamefont {Felix}\ \bibnamefont {R\"onchen}}, \bibinfo {author} {\bibfnamefont {Alec}\ \bibnamefont {Jenkins}}, \ and\ \bibinfo {author} {\bibfnamefont {Adam~M.}\ \bibnamefont {Kaufman}},\ }\bibfield  {title} {\enquote {\bibinfo {title} {Midcircuit operations using the omg architecture in neutral atom arrays},}\ }\href {\doibase 10.1103/PhysRevX.13.041035} {\bibfield  {journal} {\bibinfo  {journal} {Phys. Rev. X}\ }\textbf {\bibinfo {volume} {13}},\ \bibinfo {pages} {041035} (\bibinfo {year} {2023})}\BibitemShut {NoStop}%
\bibitem [{\citenamefont {Norcia}\ \emph {et~al.}(2023)\citenamefont {Norcia}, \citenamefont {Cairncross}, \citenamefont {Barnes}, \citenamefont {Battaglino}, \citenamefont {Brown}, \citenamefont {Brown}, \citenamefont {Cassella}, \citenamefont {Chen}, \citenamefont {Coxe}, \citenamefont {Crow}, \citenamefont {Epstein}, \citenamefont {Griger}, \citenamefont {Jones}, \citenamefont {Kim}, \citenamefont {Kindem}, \citenamefont {King}, \citenamefont {Kondov}, \citenamefont {Kotru}, \citenamefont {Lauigan}, \citenamefont {Li}, \citenamefont {Lu}, \citenamefont {Megidish}, \citenamefont {Marjanovic}, \citenamefont {McDonald}, \citenamefont {Mittiga}, \citenamefont {Muniz}, \citenamefont {Narayanaswami}, \citenamefont {Nishiguchi}, \citenamefont {Notermans}, \citenamefont {Paule}, \citenamefont {Pawlak}, \citenamefont {Peng}, \citenamefont {Ryou}, \citenamefont {Smull}, \citenamefont {Stack}, \citenamefont {Stone}, \citenamefont {Sucich}, \citenamefont {Urbanek}, \citenamefont {van~de Veerdonk}, \citenamefont
  {Vendeiro}, \citenamefont {Wilkason}, \citenamefont {Wu}, \citenamefont {Xie}, \citenamefont {Zhang},\ and\ \citenamefont {Bloom}}]{Norcia2023Midcircuit}%
  \BibitemOpen
  \bibfield  {author} {\bibinfo {author} {\bibfnamefont {M.~A.}\ \bibnamefont {Norcia}}, \bibinfo {author} {\bibfnamefont {W.~B.}\ \bibnamefont {Cairncross}}, \bibinfo {author} {\bibfnamefont {K.}~\bibnamefont {Barnes}}, \bibinfo {author} {\bibfnamefont {P.}~\bibnamefont {Battaglino}}, \bibinfo {author} {\bibfnamefont {A.}~\bibnamefont {Brown}}, \bibinfo {author} {\bibfnamefont {M.~O.}\ \bibnamefont {Brown}}, \bibinfo {author} {\bibfnamefont {K.}~\bibnamefont {Cassella}}, \bibinfo {author} {\bibfnamefont {C.-A.}\ \bibnamefont {Chen}}, \bibinfo {author} {\bibfnamefont {R.}~\bibnamefont {Coxe}}, \bibinfo {author} {\bibfnamefont {D.}~\bibnamefont {Crow}}, \bibinfo {author} {\bibfnamefont {J.}~\bibnamefont {Epstein}}, \bibinfo {author} {\bibfnamefont {C.}~\bibnamefont {Griger}}, \bibinfo {author} {\bibfnamefont {A.~M.~W.}\ \bibnamefont {Jones}}, \bibinfo {author} {\bibfnamefont {H.}~\bibnamefont {Kim}}, \bibinfo {author} {\bibfnamefont {J.~M.}\ \bibnamefont {Kindem}}, \bibinfo {author} {\bibfnamefont
  {J.}~\bibnamefont {King}}, \bibinfo {author} {\bibfnamefont {S.~S.}\ \bibnamefont {Kondov}}, \bibinfo {author} {\bibfnamefont {K.}~\bibnamefont {Kotru}}, \bibinfo {author} {\bibfnamefont {J.}~\bibnamefont {Lauigan}}, \bibinfo {author} {\bibfnamefont {M.}~\bibnamefont {Li}}, \bibinfo {author} {\bibfnamefont {M.}~\bibnamefont {Lu}}, \bibinfo {author} {\bibfnamefont {E.}~\bibnamefont {Megidish}}, \bibinfo {author} {\bibfnamefont {J.}~\bibnamefont {Marjanovic}}, \bibinfo {author} {\bibfnamefont {M.}~\bibnamefont {McDonald}}, \bibinfo {author} {\bibfnamefont {T.}~\bibnamefont {Mittiga}}, \bibinfo {author} {\bibfnamefont {J.~A.}\ \bibnamefont {Muniz}}, \bibinfo {author} {\bibfnamefont {S.}~\bibnamefont {Narayanaswami}}, \bibinfo {author} {\bibfnamefont {C.}~\bibnamefont {Nishiguchi}}, \bibinfo {author} {\bibfnamefont {R.}~\bibnamefont {Notermans}}, \bibinfo {author} {\bibfnamefont {T.}~\bibnamefont {Paule}}, \bibinfo {author} {\bibfnamefont {K.~A.}\ \bibnamefont {Pawlak}}, \bibinfo {author} {\bibfnamefont
  {L.~S.}\ \bibnamefont {Peng}}, \bibinfo {author} {\bibfnamefont {A.}~\bibnamefont {Ryou}}, \bibinfo {author} {\bibfnamefont {A.}~\bibnamefont {Smull}}, \bibinfo {author} {\bibfnamefont {D.}~\bibnamefont {Stack}}, \bibinfo {author} {\bibfnamefont {M.}~\bibnamefont {Stone}}, \bibinfo {author} {\bibfnamefont {A.}~\bibnamefont {Sucich}}, \bibinfo {author} {\bibfnamefont {M.}~\bibnamefont {Urbanek}}, \bibinfo {author} {\bibfnamefont {R.~J.~M.}\ \bibnamefont {van~de Veerdonk}}, \bibinfo {author} {\bibfnamefont {Z.}~\bibnamefont {Vendeiro}}, \bibinfo {author} {\bibfnamefont {T.}~\bibnamefont {Wilkason}}, \bibinfo {author} {\bibfnamefont {T.-Y.}\ \bibnamefont {Wu}}, \bibinfo {author} {\bibfnamefont {X.}~\bibnamefont {Xie}}, \bibinfo {author} {\bibfnamefont {X.}~\bibnamefont {Zhang}}, \ and\ \bibinfo {author} {\bibfnamefont {B.~J.}\ \bibnamefont {Bloom}},\ }\bibfield  {title} {\enquote {\bibinfo {title} {Midcircuit qubit measurement and rearrangement in a $^{171}\mathrm{Yb}$ atomic array},}\ }\href {\doibase
  10.1103/PhysRevX.13.041034} {\bibfield  {journal} {\bibinfo  {journal} {Phys. Rev. X}\ }\textbf {\bibinfo {volume} {13}},\ \bibinfo {pages} {041034} (\bibinfo {year} {2023})}\BibitemShut {NoStop}%
\bibitem [{\citenamefont {Allcock}\ \emph {et~al.}(2021)\citenamefont {Allcock}, \citenamefont {Campbell}, \citenamefont {Chiaverini}, \citenamefont {Chuang}, \citenamefont {Hudson}, \citenamefont {Moore}, \citenamefont {Ransford}, \citenamefont {Roman}, \citenamefont {Sage},\ and\ \citenamefont {Wineland}}]{Allcock2021}%
  \BibitemOpen
  \bibfield  {author} {\bibinfo {author} {\bibfnamefont {D.~T.~C.}\ \bibnamefont {Allcock}}, \bibinfo {author} {\bibfnamefont {W.~C.}\ \bibnamefont {Campbell}}, \bibinfo {author} {\bibfnamefont {J.}~\bibnamefont {Chiaverini}}, \bibinfo {author} {\bibfnamefont {I.~L.}\ \bibnamefont {Chuang}}, \bibinfo {author} {\bibfnamefont {E.~R.}\ \bibnamefont {Hudson}}, \bibinfo {author} {\bibfnamefont {I.~D.}\ \bibnamefont {Moore}}, \bibinfo {author} {\bibfnamefont {A.}~\bibnamefont {Ransford}}, \bibinfo {author} {\bibfnamefont {C.}~\bibnamefont {Roman}}, \bibinfo {author} {\bibfnamefont {J.~M.}\ \bibnamefont {Sage}}, \ and\ \bibinfo {author} {\bibfnamefont {D.~J.}\ \bibnamefont {Wineland}},\ }\bibfield  {title} {\enquote {\bibinfo {title} {Omg blueprint for trapped ion quantum computing with metastable states},}\ }\href {\doibase 10.1063/5.0069544} {\bibfield  {journal} {\bibinfo  {journal} {Appl. Phys. Lett.}\ }\textbf {\bibinfo {volume} {119}},\ \bibinfo {pages} {214002} (\bibinfo {year} {2021})}\BibitemShut {NoStop}%
\bibitem [{\citenamefont {Katori}\ \emph {et~al.}(2001)\citenamefont {Katori}, \citenamefont {Ido}, \citenamefont {Isoya},\ and\ \citenamefont {Kuwata-Gonokami}}]{Katori2001AIPLaserCoolingSr}%
  \BibitemOpen
  \bibfield  {author} {\bibinfo {author} {\bibfnamefont {Hidetoshi}\ \bibnamefont {Katori}}, \bibinfo {author} {\bibfnamefont {Tetsuya}\ \bibnamefont {Ido}}, \bibinfo {author} {\bibfnamefont {Yoshitomo}\ \bibnamefont {Isoya}}, \ and\ \bibinfo {author} {\bibfnamefont {Makoto}\ \bibnamefont {Kuwata-Gonokami}},\ }\bibfield  {title} {\enquote {\bibinfo {title} {{Laser cooling of strontium atoms toward quantum degeneracy}},}\ }\href {\doibase 10.1063/1.1354362} {\bibfield  {journal} {\bibinfo  {journal} {AIP Conf. Proc.}\ }\textbf {\bibinfo {volume} {551}},\ \bibinfo {pages} {382--396} (\bibinfo {year} {2001})}\BibitemShut {NoStop}%
\bibitem [{\citenamefont {Yang}\ \emph {et~al.}(2007)\citenamefont {Yang}, \citenamefont {Halder}, \citenamefont {Appel}, \citenamefont {Hansen},\ and\ \citenamefont {Hemmerich}}]{Yang2007Ca3P2Loading}%
  \BibitemOpen
  \bibfield  {author} {\bibinfo {author} {\bibfnamefont {C.~Y.}\ \bibnamefont {Yang}}, \bibinfo {author} {\bibfnamefont {P.}~\bibnamefont {Halder}}, \bibinfo {author} {\bibfnamefont {O.}~\bibnamefont {Appel}}, \bibinfo {author} {\bibfnamefont {D.}~\bibnamefont {Hansen}}, \ and\ \bibinfo {author} {\bibfnamefont {A.}~\bibnamefont {Hemmerich}},\ }\bibfield  {title} {\enquote {\bibinfo {title} {Continuous loading of $^{1}\mathit{S}_{0}$ calcium atoms into an optical dipole trap},}\ }\href {\doibase 10.1103/PhysRevA.76.033418} {\bibfield  {journal} {\bibinfo  {journal} {Phys. Rev. A}\ }\textbf {\bibinfo {volume} {76}},\ \bibinfo {pages} {033418} (\bibinfo {year} {2007})}\BibitemShut {NoStop}%
\bibitem [{\citenamefont {Chen}\ \emph {et~al.}(2023)\citenamefont {Chen}, \citenamefont {Bennetts},\ and\ \citenamefont {Schreck}}]{Chen2023CWBECreview}%
  \BibitemOpen
  \bibfield  {author} {\bibinfo {author} {\bibfnamefont {Chun-Chia}\ \bibnamefont {Chen}}, \bibinfo {author} {\bibfnamefont {Shayne}\ \bibnamefont {Bennetts}}, \ and\ \bibinfo {author} {\bibfnamefont {Florian}\ \bibnamefont {Schreck}},\ }\bibfield  {title} {\enquote {\bibinfo {title} {Chapter {S}ix - {T}he path to continuous {B}ose–{E}instein condensation},}\ }in\ \href {https://www.sciencedirect.com/science/article/pii/S1049250X23000046} {\emph {\bibinfo {booktitle} {Advances In Atomic, Molecular, and Optical Physics}}},\ Vol.~\bibinfo {volume} {72}\ (\bibinfo  {publisher} {Academic Press},\ \bibinfo {year} {2023})\ pp.\ \bibinfo {pages} {361--430}\BibitemShut {NoStop}%
\bibitem [{\citenamefont {Takeuchi}\ \emph {et~al.}(2023)\citenamefont {Takeuchi}, \citenamefont {Chiba}, \citenamefont {Okaba}, \citenamefont {Takamoto}, \citenamefont {Tsuji},\ and\ \citenamefont {Katori}}]{Takeuchi2023CWSrBeam}%
  \BibitemOpen
  \bibfield  {author} {\bibinfo {author} {\bibfnamefont {Ryoto}\ \bibnamefont {Takeuchi}}, \bibinfo {author} {\bibfnamefont {Hayaki}\ \bibnamefont {Chiba}}, \bibinfo {author} {\bibfnamefont {Shoichi}\ \bibnamefont {Okaba}}, \bibinfo {author} {\bibfnamefont {Masao}\ \bibnamefont {Takamoto}}, \bibinfo {author} {\bibfnamefont {Shigenori}\ \bibnamefont {Tsuji}}, \ and\ \bibinfo {author} {\bibfnamefont {Hidetoshi}\ \bibnamefont {Katori}},\ }\bibfield  {title} {\enquote {\bibinfo {title} {Continuous outcoupling of ultracold strontium atoms combining three different traps},}\ }\href {\doibase 10.35848/1882-0786/accb3c} {\bibfield  {journal} {\bibinfo  {journal} {Appl. Phys. Express}\ }\textbf {\bibinfo {volume} {16}},\ \bibinfo {pages} {042003} (\bibinfo {year} {2023})}\BibitemShut {NoStop}%
\bibitem [{\citenamefont {Hobson}\ \emph {et~al.}(2020)\citenamefont {Hobson}, \citenamefont {Bowden}, \citenamefont {Vianello}, \citenamefont {Hill},\ and\ \citenamefont {Gill}}]{Hobson2020IRMOT}%
  \BibitemOpen
  \bibfield  {author} {\bibinfo {author} {\bibfnamefont {R.}~\bibnamefont {Hobson}}, \bibinfo {author} {\bibfnamefont {W.}~\bibnamefont {Bowden}}, \bibinfo {author} {\bibfnamefont {A.}~\bibnamefont {Vianello}}, \bibinfo {author} {\bibfnamefont {I.~R.}\ \bibnamefont {Hill}}, \ and\ \bibinfo {author} {\bibfnamefont {Patrick}\ \bibnamefont {Gill}},\ }\bibfield  {title} {\enquote {\bibinfo {title} {Midinfrared magneto-optical trap of metastable strontium for an optical lattice clock},}\ }\href {\doibase 10.1103/PhysRevA.101.013420} {\bibfield  {journal} {\bibinfo  {journal} {Phys. Rev. A}\ }\textbf {\bibinfo {volume} {101}},\ \bibinfo {pages} {013420} (\bibinfo {year} {2020})}\BibitemShut {NoStop}%
\bibitem [{\citenamefont {Akatsuka}\ \emph {et~al.}(2021)\citenamefont {Akatsuka}, \citenamefont {Hashiguchi}, \citenamefont {Takahashi}, \citenamefont {Ohmae}, \citenamefont {Takamoto},\ and\ \citenamefont {Katori}}]{Akatsuka2021_3stageCooling}%
  \BibitemOpen
  \bibfield  {author} {\bibinfo {author} {\bibfnamefont {Tomoya}\ \bibnamefont {Akatsuka}}, \bibinfo {author} {\bibfnamefont {Koji}\ \bibnamefont {Hashiguchi}}, \bibinfo {author} {\bibfnamefont {Tadahiro}\ \bibnamefont {Takahashi}}, \bibinfo {author} {\bibfnamefont {Noriaki}\ \bibnamefont {Ohmae}}, \bibinfo {author} {\bibfnamefont {Masao}\ \bibnamefont {Takamoto}}, \ and\ \bibinfo {author} {\bibfnamefont {Hidetoshi}\ \bibnamefont {Katori}},\ }\bibfield  {title} {\enquote {\bibinfo {title} {Three-stage laser cooling of sr atoms using the $5s5p^{3}\mathit{P}_{2}$ metastable state below doppler temperatures},}\ }\href {\doibase 10.1103/PhysRevA.103.023331} {\bibfield  {journal} {\bibinfo  {journal} {Phys. Rev. A}\ }\textbf {\bibinfo {volume} {103}},\ \bibinfo {pages} {023331} (\bibinfo {year} {2021})}\BibitemShut {NoStop}%
\bibitem [{\citenamefont {Yasuda}\ and\ \citenamefont {Katori}(2004)}]{Yasuda2004Sr3P2}%
  \BibitemOpen
  \bibfield  {author} {\bibinfo {author} {\bibfnamefont {Masami}\ \bibnamefont {Yasuda}}\ and\ \bibinfo {author} {\bibfnamefont {Hidetoshi}\ \bibnamefont {Katori}},\ }\bibfield  {title} {\enquote {\bibinfo {title} {Lifetime measurement of the $^{3}\mathit{P}_{2}$ metastable state of strontium atoms},}\ }\href {\doibase 10.1103/PhysRevLett.92.153004} {\bibfield  {journal} {\bibinfo  {journal} {Phys. Rev. Lett.}\ }\textbf {\bibinfo {volume} {92}},\ \bibinfo {pages} {153004} (\bibinfo {year} {2004})}\BibitemShut {NoStop}%
\bibitem [{\citenamefont {Gr{\"u}nert}\ and\ \citenamefont {Hemmerich}(2002)}]{Hemmerich2002Ca3P2}%
  \BibitemOpen
  \bibfield  {author} {\bibinfo {author} {\bibfnamefont {Jan}\ \bibnamefont {Gr{\"u}nert}}\ and\ \bibinfo {author} {\bibfnamefont {Andreas}\ \bibnamefont {Hemmerich}},\ }\bibfield  {title} {\enquote {\bibinfo {title} {Sub-doppler magneto-optical trap for calcium},}\ }\href {\doibase 10.1103/PhysRevA.65.041401} {\bibfield  {journal} {\bibinfo  {journal} {Phys. Rev. A}\ }\textbf {\bibinfo {volume} {65}},\ \bibinfo {pages} {041401} (\bibinfo {year} {2002})}\BibitemShut {NoStop}%
\bibitem [{\citenamefont {Chen}\ \emph {et~al.}(2019{\natexlab{b}})\citenamefont {Chen}, \citenamefont {Bennetts}, \citenamefont {Gonz\'alez~Escudero}, \citenamefont {Schreck},\ and\ \citenamefont {Pasquiou}}]{Chen2019SOLD}%
  \BibitemOpen
  \bibfield  {author} {\bibinfo {author} {\bibfnamefont {Chun-Chia}\ \bibnamefont {Chen}}, \bibinfo {author} {\bibfnamefont {Shayne}\ \bibnamefont {Bennetts}}, \bibinfo {author} {\bibfnamefont {Rodrigo}\ \bibnamefont {Gonz\'alez~Escudero}}, \bibinfo {author} {\bibfnamefont {Florian}\ \bibnamefont {Schreck}}, \ and\ \bibinfo {author} {\bibfnamefont {Benjamin}\ \bibnamefont {Pasquiou}},\ }\bibfield  {title} {\enquote {\bibinfo {title} {Sisyphus optical lattice decelerator},}\ }\href {\doibase 10.1103/PhysRevA.100.023401} {\bibfield  {journal} {\bibinfo  {journal} {Phys. Rev. A}\ }\textbf {\bibinfo {volume} {100}},\ \bibinfo {pages} {023401} (\bibinfo {year} {2019}{\natexlab{b}})}\BibitemShut {NoStop}%
\bibitem [{\citenamefont {Cooper}\ \emph {et~al.}(2018)\citenamefont {Cooper}, \citenamefont {Covey}, \citenamefont {Madjarov}, \citenamefont {Porsev}, \citenamefont {Safronova},\ and\ \citenamefont {Endres}}]{Cooper2018_SrTweezer}%
  \BibitemOpen
  \bibfield  {author} {\bibinfo {author} {\bibfnamefont {Alexandre}\ \bibnamefont {Cooper}}, \bibinfo {author} {\bibfnamefont {Jacob~P.}\ \bibnamefont {Covey}}, \bibinfo {author} {\bibfnamefont {Ivaylo~S.}\ \bibnamefont {Madjarov}}, \bibinfo {author} {\bibfnamefont {Sergey~G.}\ \bibnamefont {Porsev}}, \bibinfo {author} {\bibfnamefont {Marianna~S.}\ \bibnamefont {Safronova}}, \ and\ \bibinfo {author} {\bibfnamefont {Manuel}\ \bibnamefont {Endres}},\ }\bibfield  {title} {\enquote {\bibinfo {title} {Alkaline-earth atoms in optical tweezers},}\ }\href {\doibase 10.1103/PhysRevX.8.041055} {\bibfield  {journal} {\bibinfo  {journal} {Phys. Rev. X}\ }\textbf {\bibinfo {volume} {8}},\ \bibinfo {pages} {041055} (\bibinfo {year} {2018})}\BibitemShut {NoStop}%
\bibitem [{\citenamefont {Chen}\ \emph {et~al.}(2024)\citenamefont {Chen}, \citenamefont {Siegel}, \citenamefont {Hunt}, \citenamefont {Grogan}, \citenamefont {Hassan}, \citenamefont {Beloy}, \citenamefont {Gibble}, \citenamefont {Brown},\ and\ \citenamefont {Ludlow}}]{Chen2024ClockSisyphus}%
  \BibitemOpen
  \bibfield  {author} {\bibinfo {author} {\bibfnamefont {Chun-Chia}\ \bibnamefont {Chen}}, \bibinfo {author} {\bibfnamefont {Jacob~L.}\ \bibnamefont {Siegel}}, \bibinfo {author} {\bibfnamefont {Benjamin~D.}\ \bibnamefont {Hunt}}, \bibinfo {author} {\bibfnamefont {Tanner}\ \bibnamefont {Grogan}}, \bibinfo {author} {\bibfnamefont {Youssef~S.}\ \bibnamefont {Hassan}}, \bibinfo {author} {\bibfnamefont {Kyle}\ \bibnamefont {Beloy}}, \bibinfo {author} {\bibfnamefont {Kurt}\ \bibnamefont {Gibble}}, \bibinfo {author} {\bibfnamefont {Roger~C.}\ \bibnamefont {Brown}}, \ and\ \bibinfo {author} {\bibfnamefont {Andrew~D.}\ \bibnamefont {Ludlow}},\ }\bibfield  {title} {\enquote {\bibinfo {title} {Clock-line-mediated sisyphus cooling},}\ }\href {\doibase 10.1103/PhysRevLett.133.053401} {\bibfield  {journal} {\bibinfo  {journal} {Phys. Rev. Lett.}\ }\textbf {\bibinfo {volume} {133}},\ \bibinfo {pages} {053401} (\bibinfo {year} {2024})}\BibitemShut {NoStop}%
\bibitem [{\citenamefont {Hunter}\ \emph {et~al.}(1986)\citenamefont {Hunter}, \citenamefont {Walker},\ and\ \citenamefont {Weiss}}]{Hunter1986Sr1D2}%
  \BibitemOpen
  \bibfield  {author} {\bibinfo {author} {\bibfnamefont {L.~R.}\ \bibnamefont {Hunter}}, \bibinfo {author} {\bibfnamefont {W.~A.}\ \bibnamefont {Walker}}, \ and\ \bibinfo {author} {\bibfnamefont {D.~S.}\ \bibnamefont {Weiss}},\ }\bibfield  {title} {\enquote {\bibinfo {title} {Observation of an atomic stark--electric-quadrupole interference},}\ }\href {\doibase 10.1103/PhysRevLett.56.823} {\bibfield  {journal} {\bibinfo  {journal} {Phys. Rev. Lett.}\ }\textbf {\bibinfo {volume} {56}},\ \bibinfo {pages} {823--826} (\bibinfo {year} {1986})}\BibitemShut {NoStop}%
\bibitem [{\citenamefont {Sansonetti}\ and\ \citenamefont {Nave}(2010)}]{Sansonetti2010SrLines}%
  \BibitemOpen
  \bibfield  {author} {\bibinfo {author} {\bibfnamefont {J.~E.}\ \bibnamefont {Sansonetti}}\ and\ \bibinfo {author} {\bibfnamefont {G.}~\bibnamefont {Nave}},\ }\bibfield  {title} {\enquote {\bibinfo {title} {{Wavelengths, Transition Probabilities, and Energy Levels for the Spectrum of Neutral Strontium (SrI)}},}\ }\href {\doibase 10.1063/1.3449176} {\bibfield  {journal} {\bibinfo  {journal} {J. Phys. Chem. Ref. Data}\ }\textbf {\bibinfo {volume} {39}},\ \bibinfo {pages} {033103} (\bibinfo {year} {2010})}\BibitemShut {NoStop}%
\bibitem [{\citenamefont {Butcher}(2008)}]{Butcher2008RK}%
  \BibitemOpen
  \bibfield  {author} {\bibinfo {author} {\bibfnamefont {John~Charles}\ \bibnamefont {Butcher}},\ }\bibfield  {title} {\enquote {\bibinfo {title} {Numerical methods for ordinary differential equations},}\ }\href@noop {} {\  (\bibinfo {year} {2008})}\BibitemShut {NoStop}%
\bibitem [{\citenamefont {Berglund}\ \emph {et~al.}(2008)\citenamefont {Berglund}, \citenamefont {Hanssen},\ and\ \citenamefont {McClelland}}]{McClelland2008CrNarrowline}%
  \BibitemOpen
  \bibfield  {author} {\bibinfo {author} {\bibfnamefont {Andrew~J.}\ \bibnamefont {Berglund}}, \bibinfo {author} {\bibfnamefont {James~L.}\ \bibnamefont {Hanssen}}, \ and\ \bibinfo {author} {\bibfnamefont {Jabez~J.}\ \bibnamefont {McClelland}},\ }\bibfield  {title} {\enquote {\bibinfo {title} {Narrow-line magneto-optical cooling and trapping of strongly magnetic atoms},}\ }\href {\doibase 10.1103/PhysRevLett.100.113002} {\bibfield  {journal} {\bibinfo  {journal} {Phys. Rev. Lett.}\ }\textbf {\bibinfo {volume} {100}},\ \bibinfo {pages} {113002} (\bibinfo {year} {2008})}\BibitemShut {NoStop}%
\bibitem [{\citenamefont {Chen}(2009)}]{chen2009active}%
  \BibitemOpen
  \bibfield  {author} {\bibinfo {author} {\bibfnamefont {JingBiao}\ \bibnamefont {Chen}},\ }\bibfield  {title} {\enquote {\bibinfo {title} {Active optical clock},}\ }\href@noop {} {\bibfield  {journal} {\bibinfo  {journal} {Chin. Sci. Bull.}\ }\textbf {\bibinfo {volume} {54}},\ \bibinfo {pages} {348--352} (\bibinfo {year} {2009})}\BibitemShut {NoStop}%
\bibitem [{\citenamefont {Meiser}\ \emph {et~al.}(2009)\citenamefont {Meiser}, \citenamefont {Ye}, \citenamefont {Carlson},\ and\ \citenamefont {Holland}}]{meiser2009prospects}%
  \BibitemOpen
  \bibfield  {author} {\bibinfo {author} {\bibfnamefont {D.}~\bibnamefont {Meiser}}, \bibinfo {author} {\bibfnamefont {Jun}\ \bibnamefont {Ye}}, \bibinfo {author} {\bibfnamefont {D.~R.}\ \bibnamefont {Carlson}}, \ and\ \bibinfo {author} {\bibfnamefont {M.~J.}\ \bibnamefont {Holland}},\ }\bibfield  {title} {\enquote {\bibinfo {title} {Prospects for a millihertz-linewidth laser},}\ }\href {\doibase 10.1103/PhysRevLett.102.163601} {\bibfield  {journal} {\bibinfo  {journal} {Phys. Rev. Lett.}\ }\textbf {\bibinfo {volume} {102}},\ \bibinfo {pages} {163601} (\bibinfo {year} {2009})}\BibitemShut {NoStop}%
\bibitem [{\citenamefont {Norcia}\ \emph {et~al.}(2016)\citenamefont {Norcia}, \citenamefont {Winchester}, \citenamefont {Cline},\ and\ \citenamefont {Thompson}}]{Norcia2016superradiance}%
  \BibitemOpen
  \bibfield  {author} {\bibinfo {author} {\bibfnamefont {Matthew~A.}\ \bibnamefont {Norcia}}, \bibinfo {author} {\bibfnamefont {Matthew~N.}\ \bibnamefont {Winchester}}, \bibinfo {author} {\bibfnamefont {Julia R.~K.}\ \bibnamefont {Cline}}, \ and\ \bibinfo {author} {\bibfnamefont {James~K.}\ \bibnamefont {Thompson}},\ }\bibfield  {title} {\enquote {\bibinfo {title} {Superradiance on the millihertz linewidth strontium clock transition},}\ }\href {\doibase 10.1126/sciadv.1601231} {\bibfield  {journal} {\bibinfo  {journal} {Sci. Adv.}\ }\textbf {\bibinfo {volume} {2}},\ \bibinfo {pages} {e1601231} (\bibinfo {year} {2016})}\BibitemShut {NoStop}%
\bibitem [{\citenamefont {Kristensen}\ \emph {et~al.}(2023)\citenamefont {Kristensen}, \citenamefont {Bohr}, \citenamefont {Robinson-Tait}, \citenamefont {Zelevinsky}, \citenamefont {Thomsen},\ and\ \citenamefont {M\"uller}}]{kristensen2023subnatural}%
  \BibitemOpen
  \bibfield  {author} {\bibinfo {author} {\bibfnamefont {Sofus~Laguna}\ \bibnamefont {Kristensen}}, \bibinfo {author} {\bibfnamefont {Eliot}\ \bibnamefont {Bohr}}, \bibinfo {author} {\bibfnamefont {Julian}\ \bibnamefont {Robinson-Tait}}, \bibinfo {author} {\bibfnamefont {Tanya}\ \bibnamefont {Zelevinsky}}, \bibinfo {author} {\bibfnamefont {Jan~W.}\ \bibnamefont {Thomsen}}, \ and\ \bibinfo {author} {\bibfnamefont {J\"org~Helge}\ \bibnamefont {M\"uller}},\ }\bibfield  {title} {\enquote {\bibinfo {title} {Subnatural linewidth superradiant lasing with cold $^{88}\mathrm{Sr}$ atoms},}\ }\href {\doibase 10.1103/PhysRevLett.130.223402} {\bibfield  {journal} {\bibinfo  {journal} {Phys. Rev. Lett.}\ }\textbf {\bibinfo {volume} {130}},\ \bibinfo {pages} {223402} (\bibinfo {year} {2023})}\BibitemShut {NoStop}%
\bibitem [{dat()}]{dataavailability}%
  \BibitemOpen
  \href {\doibase https://doi.org/10.5281/zenodo.14964616} {\ https://doi.org/10.5281/zenodo.14964616}\BibitemShut {NoStop}%
\end{thebibliography}
%

\end{document}